\documentclass[11pt]{book}

\usepackage[dvips]{epsfig}
\usepackage{epsf}
\usepackage{amssymb}
\usepackage{multirow}
\usepackage{amsbsy}
\usepackage{pstricks}
\catcode`\@=11

 \def\@normalsize{\@setsize\normalsize{13pt}\xipt\@xipt
   \abovedisplayskip 11pt plus3pt minus6pt
   \belowdisplayskip \abovedisplayskip
   \abovedisplayshortskip \z@ plus3pt
   \belowdisplayshortskip 6.6pt plus3.5pt minus3pt}

 \def\small{\@setsize\small{12pt}\xipt\@xipt
   \abovedisplayskip 10pt plus2pt minus5pt
   \belowdisplayskip \abovedisplayskip
   \abovedisplayshortskip \z@ plus3pt
   \belowdisplayshortskip 6pt plus3pt minus3pt
   \def\@listi{\topsep 6pt plus 2pt minus 2pt
     \parsep 3pt plus 2pt minus 1pt
     \itemsep \parsep}}

 \def\footnotesize{\@setsize\footnotesize{10pt}\ixpt\@ixpt
   \abovedisplayskip 8pt plus 2pt minus 4pt
   \belowdisplayskip \abovedisplayskip
   \abovedisplayshortskip \z@ plus 1pt
   \belowdisplayshortskip 4pt plus 2pt minus 2pt
   \def\@listi{\topsep 4pt plus 2pt minus 2pt
      \parsep 2pt plus 1pt minus 1pt
      \itemsep \parsep}}

 \def\scriptsize{\@setsize\scriptsize{9.5pt}\viiipt\@viiipt}
 \def\tiny{\@setsize\tiny{7pt}\vipt\@vipt}
 \def\large{\@setsize\large{14pt}\xiipt\@xiipt}
 \def\Large{\@setsize\Large{18pt}\xivpt\@xivpt}
 \def\LARGE{\@setsize\LARGE{22pt}\xviipt\@xviipt}
 \def\huge{\@setsize\huge{25pt}\xxpt\@xxpt}
 \def\Huge{\@setsize\Huge{30pt}\xxvpt\@xxvpt}
\def\section{\@startsection {section}{1}{\z@}%
{-1.5\baselineskip plus-1pt minus-3pt}{1\baselineskip plus1pt minus2pt}%
{\centering\normalsize\bf}}
\def\subsection{\@startsection{subsection}{2}{\z@}%
{-1\baselineskip plus-1pt minus-2pt}{1\baselineskip plus1pt minus2pt}%
{\normalsize\sc\noindent}}
\def\subsubsection{\@startsection{subsubsection}{3}{\z@}%
{-1\baselineskip plus-1pt minus-2pt}{1sp}{\normalsize\it\noindent}}
\def\paragraph{\@startsection{paragraph}{4}{\z@}%
{1\baselineskip plus1pt minus2pt}{-1em}{\normalsize\it\noindent}}
\let\subparagraph=\paragraph

\def\tableofcontents{\@restonecolfalse\if@twocolumn\@restonecoltrue
\onecolumn\fi\OSIDcont\@starttoc{con}\if@restonecol\twocolumn\fi}

\def\l@section{\@dottedtocline{1}{0em}{.66em}}

\def\thebibliography#1{\section*{{Bibliography}\@mkboth
 {BIBLIOGRAPHY}{BIBLIOGRAPHY}}\footnotesize\rm\list
 {[\arabic{enumi}]}{\settowidth\labelwidth{[#1]}\leftmargin\labelwidth
 \advance\leftmargin\labelsep\usecounter{enumi}}
 \def\newblock{\hskip .11em plus .33em minus -.07em}
 \sloppy\clubpenalty4000\widowpenalty4000
 \sfcode`\.=1000\relax}

\def\ps@myheadings{\let\@mkboth\@gobbletwo
\def\@oddhead{\hfil{\footnotesize\rm\rightmark}\hfil}
\def\@evenhead{\hfil{\footnotesize\rm\leftmark}\hfil}
\def\@oddfoot{\hfil{\footnotesize\sf\artid-\thepage}\hfil}
\def\@evenfoot{\hfil{\footnotesize\sf\artid-\thepage}\hfil}
\def\sectionmark##1{}\def\subsectionmark##1{}}

\def\@copyrighthead{\parbox{127mm}{\footnotesize\rm\ \\[6pt]
Open Systems~\& Information Dynamics\\
Vol.~\Vol, No.~\Number~(\Year)~\artid~(\EndpagE~pages)\\
DOI:\DOInumber\\
\copyright~World Scientific Publishing Company\\
\epsfxsize=4cm
\vskip-\lastskip
\vskip-\baselineskip
\vspace*{-38.5pt}
\noindent\hfill\epsfbox{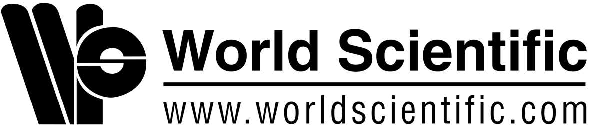}}}

\def\artid{0000001}
\def\Year{2026}        %
\def\Vol{33}           % <-------( ustawienia robocze )
\newcounter{paPer}     %
\def\EndpagE{\expandafter\pageref{\the\value{paPer}OpSy}}

\def\ps@osiD{\let\@mkboth\@gobbletwo
\def\@oddhead{\@copyrighthead}
  \def\@oddfoot{\hfil{\footnotesize\sf\artid-\thepage}\hfil}
  \def\@evenhead{}\let\@evenfoot\@oddfoot}

\def\cite{\@ifnextchar [{\@tempswatrue\@Rcitex}{\@tempswafalse\@Rcitex[]}}

\def\@Rcitex[#1]#2{\if@filesw\immediate\write\@auxout{\string\citation{#2}}\fi
  \def\@citea{}\@cite{\@for\@citeb:=#2\do
    {\@citea\def\@citea{,\penalty\@m\,}\@ifundefined
       {b@\the\value{paPer}R\@citeb}{{\bf ?}\@warning
       {Citation `\@citeb' on page \thepage \space undefined}}%
\hbox{\csname b@\the\value{paPer}R\@citeb\endcsname}}}{#1}}

\long\def\@caption#1[#2]#3{\par\addcontentsline{\csname
  ext@#1\endcsname}{#1}{\protect\numberline{\csname
  the#1\endcsname}{\ignorespaces #2}}\begingroup
    \@parboxrestore
    \small                                        %    \normalsize
    \@makecaption{\csname fnum@#1\endcsname}{\ignorespaces #3}\par
  \endgroup}

\catcode`\@=12
\let\Rlabel=\label
\let\Rbibitem=\bibitem
\let\Rref=\ref
\let\Rpageref=\pageref
\def\label#1{\expandafter\Rlabel{\the\value{paPer}R#1}}
\def\bibitem#1{\expandafter\Rbibitem{\the\value{paPer}R#1}}
\def\ref#1{\expandafter\Rref{\the\value{paPer}R#1}}
\def\pageref#1{\expandafter\Rpageref{\the\value{paPer}R#1}}

\def\thesection{\arabic{section}.}

\def\YYMm{\rule{0ex}{4em}}
\newtoks\TITsi
\newtoks\TITsii

\def\title#1{\def\TITs{\LARGE{\raggedright\noindent\YYMm #1%
\vskip8pt\par}}}
\def\author#1{\autMM{#1}\def\LHD{#1}}
\def\and{{\rm\lowercase{and}}}

\def\autMM#1{\TITsii={\vskip10pt\par\normalsize\rm\noindent #1\par}%
\TITsi=\expandafter{\TITs}\edef\TITs{\the\TITsi\the\TITsii}}

\def\address#1{\TITsii={\vskip6pt\par\footnotesize\sl\noindent #1\par}%
\TITsi=\expandafter{\TITs}%
\edef\TITs{\the\TITsi\the\TITsii}}

\def\received#1{\TITsii={\vskip10pt\par\small\rm\noindent(Received: #1)\par}%
\TITsi=\expandafter{\TITs}\edef\TITs{\the\TITsi\the\TITsii}}

\def\headtitle#1{\def\RHD{#1}}
\def\headauthor#1{\def\LHD{#1}}
\def\listas#1#2{\addcontentsline{con}{section}{{\sc #1: }{\rm #2}}}

\def\abst{{\bf Abstract.}}
\def\abstract#1{\TITs
       \vskip15pt\par\noindent
       {\footnotesize{\abst~} #1\vskip3pt\par}
       \markright{\RHD}
       \markboth{\LHD}{\RHD}}

\def\startpaper{%
       \cleardoublepage
       \setcounter{section}{0}
       \stepcounter{paPer}
       \setcounter{equation}{0}
       \setcounter{footnote}{0}
       \setcounter{figure}{0}
       \setcounter{table}{0}
       \def\theequation{\arabic{equation}}
       \def\thefootnote{\arabic{footnote}}
       \setcounter{defn}{0}
       \setcounter{thm}{0}
       \setcounter{lem}{0}
       \setcounter{prop}{0}
       \setcounter{rem}{0}
       \thispagestyle{osiD}}

\def\OSIDcont{\cleardoublepage\thispagestyle{empty}
       \markright{}\markboth{}{}
       \normalsize\rm
       \hspace*{\fill}{\large\rm
         Contents of the Volume \Volume, Number \Number}\hspace*{\fill}
       \par\vspace{1.5em}
       \par\noindent}

\def\endpaper{\expandafter\label{\the\value{paPer}OpSy}}

\def\1{{\mathchoice{\rm 1\mskip-4mu l}{\rm 1\mskip-4mu l}%
{\rm 1\mskip-4.5mu l}{\rm 1\mskip-5mu l}}}
\def\varkappa{\mbox{\bBB\char 123}}
\def\longhookrightarrow{\lhook\joinrel\relbar\joinrel\rightarrow}

\def\longhookUp{\lower6pt\hbox{\rotatebox{90}{$\longhookrightarrow$}}}

\def\theequation{\thesection\arabic{equation}}

\def\Myskip{\setlength{\baselineskip}{13pt}}
\def\text#1{\quad\mbox{\rm  #1 }\quad}

\def\DOInumber{}
\input xy
\xyoption{all}

\InputIfFileExists{psfig.sty}{\typeout{^^Jpsfig.sty inputed...ok}}{\typeout{^^JWarning: psfig.sty could not be found.^^J}}
\usepackage{graphicx}
\usepackage{amsmath,amssymb,amsfonts,mathtools,bm}%
\usepackage{bbm}
\usepackage{amsthm}%
\usepackage{mathrsfs}%
\usepackage[title]{appendix}%
\usepackage{xcolor}%
\usepackage{textcomp}%
\usepackage{manyfoot}%
\usepackage{booktabs}%
\usepackage{algorithm}%
\usepackage{algorithmicx}%
\usepackage{algpseudocode}%
\usepackage{listings}%
\usepackage{hyperref}

\newcommand{\ket}[1]{\lvert #1\rangle}
\newcommand{\bra}[1]{\langle #1\lvert}
\renewcommand{\vec}[1]{\mathbf{ #1}}
\newcommand{\opvec}[1]{\hat{\mathbf{ #1}}}
\newcommand{\kick}{\boldsymbol{\Delta}_p}
\newcommand{\cq}[1]{C_{Q}(#1)}
\newcommand{\trans}{\mathrm T}
\newcommand{\Reals}{\mathbb{R}}
\newcommand{\Gen}[1]{\mathcal{L}_{\mathrm{#1}}}
\newcommand{\pref}{\vec{p}_\mathrm{ref}}
\newcommand{\deltap}{\boldsymbol{\delta}\vec{p}}

\begin{document}

\def\Number{0}
\def\Volume{0}

\startpaper

\newcommand{\Mn}{M_n(\mathbb{C})}
\newcommand{\Mk}{M_k(\mathbb{C})}
\newcommand{\id}{\mbox{id}}
\newcommand{\ot}{{\,\otimes\,}}
\newcommand{{\Cd}}{{\mathbb{C}^d}}
\newcommand{\sbsigma}{{\mbox{\scriptsize \boldmath $\sigma$}}}
\newcommand{\sbalpha}{{\mbox{\scriptsize \boldmath $\alpha$}}}
\newcommand{\sbbeta}{{\mbox{\scriptsize \boldmath $\beta$}}}
\newcommand{\bsigma}{{\mbox{\boldmath $\sigma$}}}
\newcommand{\balpha}{{\mbox{\boldmath $\alpha$}}}
\newcommand{\bbeta}{{\mbox{\boldmath $\beta$}}}
\newcommand{\bmu}{{\mbox{\boldmath $\mu$}}}
\newcommand{\bnu}{{\mbox{\boldmath $\nu$}}}
\newcommand{\ba}{{\mbox{\boldmath $a$}}}
\newcommand{\bb}{{\mbox{\boldmath $b$}}}
\newcommand{\sba}{{\mbox{\scriptsize \boldmath $a$}}}
\newcommand{\MD}{\mathfrak{D}}
\newcommand{\sbb}{{\mbox{\scriptsize \boldmath $b$}}}
\newcommand{\sbmu}{{\mbox{\scriptsize \boldmath $\mu$}}}
\newcommand{\sbnu}{{\mbox{\scriptsize \boldmath $\nu$}}}
\def\oper{{\mathchoice{\rm 1\mskip-4mu l}{\rm 1\mskip-4mu l}%
{\rm 1\mskip-4.5mu l}{\rm 1\mskip-5mu l}}}
\def\<{\langle}
\def\>{\rangle}
\def\theequation{\thesection\arabic{equation}}

\title{Zeno-Constrained Formation of Relativistic Mass Shells}
\author{Ansgar Pernice}
\address{Kochel a. See, Germany}
\headauthor{Ansgar Pernice}
\headtitle{Zeno-Constrained Formation of Relativistic Mass Shells}
\received{February 12, 2026}
\listas{Ansgar Pernice}{Zeno-Constrained Formation of Relativistic Mass Shells}

\abstract{We study an extension of the quantum linear Boltzmann equation describing irreversible momentum-space dynamics of an open quantum system under strong continuous monitoring. The monitored observable is taken to be a quadratic form in an extended, purely Euclidean four-dimensional momentum space, without assuming any fixed signature at the microscopic level. In the resulting quantum Zeno regime, rapid suppression of off-constraint excursions allows for an adiabatic elimination of fast degrees of freedom. Using a Schur-complement construction, the induced second-order corrections give rise to an effective flow of the monitored quadratic form under temporal coarse graining. Under mild isotropy assumptions on the underlying momentum-mixing dynamics and an appropriate calibration condition, this flow approaches an infrared fixed point characterized by a quadratic form of Lorentzian signature. The corresponding null set defines a mass-shell-like constraint surface that governs the long-time Zeno-projected dynamics and whose isometry group matches the kinematic structure of Lorentz transformations at the effective level. Familiar relativistic features, including Maxwell–Jüttner-type stationary distributions, arise at the level of the effective infrared description as consequences of this fixed point within the extended quantum Boltzmann framework.}

\Myskip

%\pacs{03.65.Ud, 03.67.-a}

%\SMP

%============================================================
\section{Introduction}\setcounter{equation}{0}
\label{sec:introduction}
%============================================================

Relativistic kinematic structures are commonly introduced at the level of
effective theories through symmetry principles and mass--shell constraints.
In particular, Lorentz invariance is typically imposed as a kinematical input,
while irreversible dynamics is treated as a secondary, model--dependent
ingredient. In the present work, we adopt a complementary and technically
more modest viewpoint: we investigate how relativistic mass--shell structures
can arise as infrared features of an extended momentum--space dynamics within
a well--established open--system framework.

Irreversible dynamics is a common consequence of environmental coupling
under suitable coarse--graining assumptions.
Whenever a quantum system interacts with unobserved degrees of freedom and
the environment acts effectively as a memoryless reservoir, the reduced
dynamics is described by effective, typically Markovian equations that
break time--reversal invariance
\cite{davies1974markovian,BreuerPetruccioneBook}. In many physically relevant
situations---including dilute gases, scattering--dominated regimes, and
coarse--grained kinetic descriptions---such dynamics is most naturally
formulated in momentum space rather than in terms of spacetime variables
\cite{VacchiniHornberger2009}.

A paradigmatic example is provided by the quantum linear Boltzmann equation
(QLBE), which describes the irreversible momentum--space dynamics of a massive
test particle interacting with a background medium
\cite{HornbergerVacchini2008,VacchiniHornberger2009}. The QLBE is translationally
invariant, microscopically grounded in scattering theory, and widely accepted
as a physically consistent model of open quantum dynamics. It therefore serves
as a natural starting point for exploring how kinematic constraints may arise
within irreversible momentum--space evolutions, rather than being imposed
externally.
In this paper, we extend the standard QLBE framework by allowing the monitored
kinematic observable itself to become dynamical. Concretely, we consider the
continuous monitoring of a quadratic momentum--space observable
\[
\cq{\opvec{p}} = \opvec{p}^\trans Q\,\opvec{p} ,
\]
where $Q$ is a symmetric tensor defined on an extended, purely Euclidean
four--dimensional momentum space.

Continuous monitoring of quadratic and
energy--like observables, including kinetic energy and momentum--space
quantities, has been extensively studied in the context of open quantum
systems and stochastic measurement theory
\cite{ChantasriDresselJordan2013,wiseman2009quantum,JacobsSteck2006,Menskii1998} from
the viewpoint of three spatial momentum dimensions. In particular, gravitationally induced decoherence derived from GR+QFT leads, in the non-relativistic regime, to Lindblad generators quadratic in the momentum operator \cite{AnastopoulosHu2013}. This supports the structural naturalness of monitoring quadratic momentum-space observables as considered here. In this work, however, we formally extend the concept to four dimensions, treated on equal footing.
At the microscopic level, no assumption is made about the signature or special
symmetry properties of $Q$. Instead, $Q$ is treated as a dynamical object whose
effective form is shaped by the interplay of irreversible momentum transfer
and strong continuous monitoring.

A central technical ingredient of our analysis is a controlled quantum Zeno
regime
\cite{misra1977zeno,FacchiPascazio2002,Burgarth2020,schafer2014experimental},
in which rapid monitoring suppresses fast excursions away from the monitored
constraint surface. When combined with irreversible momentum--space dynamics,
the adiabatic elimination of these off--constraint degrees of freedom generates
systematic second--order corrections to the effective generator. Using a
Schur--complement construction, these corrections can be interpreted as a
renormalization of the quadratic form $Q$ itself. Such measurement--induced
backaction on observables, beyond simple state reduction, is a well--known
feature of continuous quantum measurement and monitored open--system dynamics
\cite{wiseman2009quantum,Becker2021,venuti2016adiabaticity}.

Iterated over successive coarse--graining steps, this mechanism induces a flow
in the space of quadratic forms. Conceptually, this places the present
construction in close relation to adiabatic elimination and strong--coupling
limits in open quantum systems, where effective generators emerge from the
systematic removal of fast degrees of freedom
\cite{SarandyLidar2005,AzouitSarletteRouchon2016}.
The distinguishing feature here is that the flow acts on the monitored
kinematic observable rather than on the system state.

Since only the relative structure of the monitored observable is physically
meaningful, the induced flow must be supplemented by a calibration condition
fixing its overall scale. Under this calibration and mild isotropy assumptions
on the underlying momentum--mixing dynamics, we show that the flow admits a
nontrivial infrared fixed point. Although the microscopic dynamics is defined
on a Euclidean momentum space, the effective quadratic form characterizing this
fixed point exhibits a Lorentzian signature. The corresponding null set
defines a mass--shell--like constraint surface that governs the long--time,
Zeno--projected dynamics.

The main result of this work is the existence and stability of such a
Lorentz--signature fixed point within the extended QLBE framework. Familiar
relativistic structures, including mass--shell constraints and
Maxwell--J\"uttner--type stationary distributions, arise at the level of the
effective infrared description as consequences of this fixed point, rather
than as assumptions imposed on the microscopic dynamics. The construction
therefore provides a concrete example of how relativistic kinematic features
can appear within a purely Euclidean, irreversible momentum--space model under
appropriate dynamical and measurement conditions.

The paper is organized as follows. In
Section~\ref{sec:qlbe} we review the quantum linear Boltzmann equation as a
microscopically grounded model of irreversible momentum--space dynamics
and generalize its structure to four Euclidean dimensions. In
Section~\ref{sec:monitoring} we introduce the monitoring of quadratic
observables and analyze the resulting Zeno--type dynamics.
Section~\ref{sec:renormalization} develops the induced renormalization flow of
the quadratic form and its calibration.
In Section~\ref{sec:fixedpoint} we demonstrate the existence and stability of
the Lorentz--signature infrared fixed point.
Section~\ref{sec:interpretation} discusses the physical interpretation and
stationary states.

%============================================================
\section{Irreversible Momentum--Space Dynamics and the QLBE}\setcounter{equation}{0}
\label{sec:qlbe}
%============================================================

To ground our approach in a physically established framework, we begin by
recalling the quantum linear Boltzmann equation (QLBE), which provides a
prototypical example of irreversible dynamics formulated directly in momentum
space. The QLBE describes the reduced dynamics of a massive test particle
interacting with a dilute background gas and is derived under well--controlled
assumptions from microscopic scattering theory
\cite{HornbergerVacchini2008,VacchiniHornberger2009}.

\subsection{Quantum linear Boltzmann equation}

The defining features of the QLBE are its translational invariance,
Markovianity, and complete positivity. In the absence of external potentials,
the generator commutes with spatial translations, so that momentum is the
natural variable in which the evolution takes a simple form. The resulting
master equation describes stochastic momentum transfers associated with
individual scattering events, leading to dissipation and decoherence in
momentum space within the general framework of translation--covariant open
quantum dynamics \cite{Holevo1993,Vacchini2001}.

Formally, the QLBE is a Lindblad master equation whose jump operators
implement finite momentum kicks. These encode the statistics of scattering
processes through a collision kernel determined by the differential cross
section and the momentum distribution of the background medium. The dynamics
is intrinsically irreversible: it breaks time--reversal symmetry and drives
the system toward equilibrium under appropriate conditions
\cite{BreuerPetruccioneBook}.

In its standard nonrelativistic formulation, the quantum linear Boltzmann
equation governs the reduced density operator $\rho$ of a free test particle according to
\begin{equation}
\label{eq:qlbe}
\begin{split}
\frac{d\rho}{dt}
=
-\frac{i}{\hbar}\,[H_{\mathrm{eff}} 
,\rho]&+
\int d^3\boldsymbol{\Delta}_p\,
\Big(
L(\boldsymbol{\Delta}_p)\,\rho\,L^\dagger(\boldsymbol{\Delta}_p)\\
&-
\frac{1}{2}\{L^\dagger(\boldsymbol{\Delta}_p)L(\boldsymbol{\Delta}_p),\rho\}
\Big),
\end{split}
\end{equation}
where $H_{\mathrm{eff}}$ denotes the effective Hamiltonian of the test particle,
including possible energy renormalizations due to the interaction with the gas
(e.g.\ Lamb--shift contributions)
\cite{HornbergerVacchini2008,VacchiniHornberger2009}.
The detailed structure of the Lamb--shift term and any additional scattering
degrees of freedom are not required for the present analysis.

The Lindblad operators $L(\kick)$ implement finite momentum
kicks and are given by
\begin{equation}
\label{eq:qlbe_jump}
L(\kick)
=
\exp\!\left(\frac{i}{\hbar}\kick\cdot\mathbf{x}\right)
\,\sqrt{\Gamma(\kick,\vec{p})},
\end{equation}
where $\Gamma(\kick,\vec{p})$ is a positive operator encoding
the scattering rate associated with a momentum transfer
$\kick$, determined by the differential cross section and the momentum
distribution of the background gas.

Equation~\eqref{eq:qlbe} therefore describes an irreversible,
translation--covariant dynamics in momentum space. Individual scattering
events induce stochastic momentum updates of the form
$\vec{p}\mapsto\vec{p}+\kick$, leading to dissipation and
decoherence. In its standard formulation the dynamics is nonrelativistic:
the momentum variable is three--dimensional, the kinetic energy is quadratic,
and no mass--shell constraint or Lorentzian structure is assumed.

For the purposes of the present work, the detailed form of the collision
kernel is not essential. What matters is that the QLBE provides a concrete,
microscopically motivated realization of an open--system dynamics acting
directly on momentum degrees of freedom, with a clear interpretation in
terms of stochastic momentum transfers and environmental monitoring
\cite{HornbergerVacchini2008,VacchiniHornberger2009}.

We will therefore use its structural features as a template for a more
general class of irreversible momentum--space dynamics, without modifying
its basic physical content.

\subsection{Extension to four Euclidean dimensions}
In the following, we formally extend this structure to a four--dimensional
Euclidean momentum space,
$\kick,\,\vec{p}\in\Reals^4$, equipped with the standard
Euclidean inner product $\delta_{\mu\nu}$. All four momentum components are
treated on equal footing, and no preferred direction is singled out by the
microscopic dynamics. Equation~\eqref{eq:qlbe} then defines an irreversible
generator of the form
\begin{equation}
\label{eq:general_kicks}
\Gen{irr}[\rho]
=
\int d^4\kick\,
\Big(
L(\kick)\,\rho\,L^\dagger(\kick)
-
\frac{1}{2}\{L^\dagger(\kick)L(\kick),\rho\}
\Big),
\end{equation}
which is isotropic and fully Euclidean at the fundamental level.

Within this irreversible momentum--space dynamics, we consider the continuous
monitoring of a quadratic observable of the form
\begin{equation}
\label{eq:CQ_def}
C_Q(\opvec{p}) = \opvec{p}^\trans Q\, \opvec{p} ,
\end{equation}
where $Q$ is a real, symmetric $4\times 4$ matrix.
The monitoring is modeled by an additional pure--dephasing generator,
\begin{equation}
\label{eq:monitoring}
\Gen{mon}[\rho]
=
\kappa\bigl(
C_Q \rho C_Q - \tfrac12\{C_Q^2,\rho\}
\bigr),
\end{equation}
which suppresses coherences between states with different values of $C_Q$ and
induces a Zeno--type separation of time scales for large monitoring strength
$\kappa$ \cite{misra1977zeno,FacchiPascazio2002,Burgarth2020}.
Physically, such monitoring can arise in gas or plasma environments when
scattering rates and collision statistics depend on quadratic combinations of
the momentum. Through many weak interactions, the environment acquires
information about $\vec{p}^\trans Q\,\vec{p}$, leading—after coarse graining—to effective
dephasing in the corresponding eigenbasis
\cite{wiseman2009quantum}.

\subsection{Coarse--grained evolution and relation to physical time}

Taken together, the reduced dynamics is governed by a Lindblad-type master equation
\begin{equation}
\frac{d\rho}{d\lambda}
=
-\frac{i}{\hbar}\,[H_{\mathrm{eff}},\rho]
+\Gen{irr}[\rho]+\Gen{mon}[\rho]
= \Gen{}[\rho].
\end{equation}

At this stage, the parameter $\lambda$ should be regarded purely as a coarse--grained flow parameter.
It labels successive elimination and renormalization steps within the synthetic four--dimensional momentum--space description introduced above.
No identification with physical time is assumed. The replacement $t \rightarrow \lambda$ therefore reflects a deliberate separation between microscopic time evolution and the effective flow generated by $\Gen{}$.
Similar reparametrizations appear in coarse--grained kinetic equations and in adiabatic elimination schemes in open quantum systems \cite{Davies1979,AzouitSarletteRouchon2016}.

Only at the level of the effective infrared description --- after the dynamical selection of a preferred geometric structure in momentum space --- does the question arise whether one may consistently interpret part of the reduced dynamics in terms of a physical time parameter.
Any such identification is therefore emergent and interpretative rather than fundamental within the present construction.

\paragraph*{Remark:}At this stage, no assumption is made about relativistic symmetry, mass--shell
constraints, or the existence of a preferred time direction. The quadratic
form $Q$ is not fixed \emph{a priori} and is not assumed to have a Lorentzian
signature. The purpose of introducing the four--dimensional Euclidean
extension~\eqref{eq:general_kicks}--\eqref{eq:monitoring} is therefore not to
postulate relativistic kinematics, but to investigate whether such structures
can arise dynamically from the combined effect of irreversible momentum--space
evolution and continuous monitoring.

%============================================================
\section{Monitoring of Quadratic Observables and Zeno Dynamics}\setcounter{equation}{0}
\label{sec:monitoring}
%============================================================

In this section we analyze the dynamical consequences of continuously
monitoring a quadratic observable in momentum space.
Our goal is to show that, in the strong--monitoring (Zeno) regime,
irreversible momentum--space dynamics does not merely constrain the system
to fixed level sets of the monitored quantity, but feeds back onto the
observable itself.
This feedback takes the form of an effective, measurement--induced
renormalization of the quadratic form that defines the monitored observable,
a phenomenon that goes beyond state reduction and has been discussed in
various forms in the context of monitored open quantum systems
\cite{ChantasriDresselJordan2013}.
In contrast to these works, where measurement backaction is primarily
analyzed at the level of state trajectories, conditioned dynamics, or
effective Hamiltonians for a fixed measured operator, we demonstrate that
unconditional irreversible dynamics can induce a genuine structural
renormalization of the monitored observable itself.

%------------------------------------------------------------
\subsection{Zeno regime and separation of time scales}
%------------------------------------------------------------

We consider continuous monitoring of the quadratic momentum--space observable $C_Q(\opvec{p})=\opvec{p}^\trans Q \opvec{p}$ implemented by the pure--dephasing generator $\Gen{mon}$ defined in~\eqref{eq:monitoring}.
Such generators arise naturally in the theory of continuous quantum
measurement and stochastic master equations and describe dephasing induced by
the acquisition of information about the observable $C_Q$ by the environment
\cite{wiseman2009quantum,ChantasriDresselJordan2013}.

For large monitoring strength $\kappa$, coherences between states with
different values of $C_Q$ are rapidly suppressed.
The dynamics therefore enters a Zeno regime characterized by a pronounced
separation of time scales, as originally identified in the context of
frequent measurements and strong continuous coupling and later generalized to
open quantum systems and arbitrary observables
\cite{misra1977zeno,FacchiPascazio2002,Burgarth2020,Becker2021}.

On short time scales of order $\kappa^{-1}$, the density operator becomes
approximately block--diagonal in the eigenbasis of $C_Q$.
On much longer time scales, residual dynamics survives within the
corresponding Zeno subspace.
This slow subspace consists of momentum--space coherences
$\rho(p,p')$ connecting momenta $p$ and $p'$ that lie on (or very close to)
the same level set of the monitored observable.

%------------------------------------------------------------
\subsection{Interplay with irreversible momentum--space dynamics}
%------------------------------------------------------------

The monitoring dynamics is supplemented by an irreversible momentum--space
generator $\mathcal L_{\rm irr}$ describing stochastic momentum transfers.
Such generators arise naturally in collisional models and in the quantum
linear Boltzmann equation
\cite{HornbergerVacchini2008,VacchiniHornberger2009}.
Finite momentum kicks generically change the value of $C_Q$ and are therefore
strongly suppressed by the monitoring term.

As a result, the combined dynamics exhibits a clear hierarchy:
off--shell excursions are inhibited at leading order, but contribute
virtually at higher order through sequences of suppressed transitions.
This structure is characteristic of strong--coupling and adiabatic
elimination regimes in open quantum systems, including Zeno and
singular--perturbation limits
\cite{SarandyLidar2005,AzouitSarletteRouchon2016,Becker2021,venuti2016adiabaticity}.

To analyze this structure, we decompose the full generator as
\begin{equation}
\mathcal L
=
\mathcal L_{\rm slow}
+
\mathcal L_{\rm mix}
+
\mathcal L_{\rm mon} ,
\end{equation}
where $\mathcal L_{\rm mix}$ denotes the part of the irreversible dynamics
that changes the value of $C_Q$.
Standard projection and adiabatic--elimination techniques then yield an
effective generator for the slow Zeno dynamics of the form
\begin{equation}
\label{eq:schur}
\mathcal L_{\rm eff}
=
P_Z\,\mathcal L_{\rm slow}\,P_Z
-
P_Z\,\mathcal L_{\rm mix}\,
\bigl(Q_Z\mathcal L_{\rm mon}Q_Z\bigr)^{-1}
\mathcal L_{\rm mix}\,P_Z ,
\end{equation}
where $P_Z$ projects onto the Zeno subspace and $Q_Z=1-P_Z$ onto its complement.
The second term captures virtual ``kick--out and return'' processes and is a
standard Schur--complement contribution familiar from adiabatic elimination
and singular perturbation theory in open quantum systems
\cite{davies1974markovian,AzouitSarletteRouchon2016}. Under assumptions (A1)–(A5) stated in Appendix~\ref{app:monitoring_zeno},
the reduced generator $\mathcal L_{\rm eff}$ remains of Gorini--Kossakowski--Sudarshan--Lindblad (GKSL) -- type
on the Zeno sector, inheriting complete positivity and trace preservation
from the underlying Markovian dynamics.

%------------------------------------------------------------
\subsection{Reference structure: shifting the monitored observable}
%------------------------------------------------------------

Before analyzing the Schur--complement term, it is useful to establish a
reference structure.
Consider a small shift of the quadratic form,
\begin{equation}
Q\;\longrightarrow\;Q' = Q-\delta Q ,
\qquad \|\delta Q\|\ll\|Q\| .
\end{equation}
Since the monitored observable depends linearly on $Q$, this induces a
corresponding variation of the monitoring generator.
To first order in $\delta Q$, the variation produces an additional pure
dephasing contribution of the form
\begin{equation}
\label{eq:deltaLmonRate}
\bra{\vec{p}}\delta\mathcal L_{\rm mon}[\rho]\ket{\vec{p}'}
=
\kappa\,
\Delta C_Q(\vec{p},\vec{p}')\,\Delta C_{\delta Q}(\vec{p},\vec{p}')\,
\rho(\vec{p},\vec{p}') ,
\end{equation}
where $\Delta C_Q(\vec{p},\vec{p}')=C_Q(\vec{p})-C_Q(\vec{p}')$.
This expression provides a structural template for identifying the effect of
the Schur--complement correction at a later stage and mirrors general results on the response
of monitored generators to perturbations of the measured observable
\cite{Kessler2012,ReiterSorensen2012}. At this stage, the expression is purely algebraic and does
not yet invoke any local geometric interpretation; locality
enters only after restriction to the Zeno sector being discussed in the next
subsection.

%------------------------------------------------------------
\subsection{Restriction to the Zeno sector and emergence of local geometry}
%------------------------------------------------------------

Up to this point, the description has been formulated entirely
at the level of operators on momentum space, and the quadratic
observable $C_Q(\opvec{p})$ was specified by a global
parameter $Q$ without any reference to local geometric
structures. \textbf{In particular, no notion of a reference
momentum, tangent space, or local decomposition is meaningful
at this stage.}

This viewpoint has to be reconsidered in the Zeno limit, where the monitoring dynamics suppresses coherences between momentum
eigenstates with different values of $C_Q$. As a result, the effective dynamics
is confined to near--diagonal momentum coherences $(\vec{p_1},\vec{p_2})$ with
$\vec{p_1}\approx \vec{p_2}$. From this point on, the description becomes
intrinsically local in momentum space: the relevant degrees of freedom are pairs
$(\pref,\vec{\deltap})$, and geometric notions such as normal and tangent
directions emerge through a local linearization of $C_Q$.
All geometric constructions introduced below are therefore understood as local
objects defined at a reference momentum $\pref$.

%------------------------------------------------------------
\subsection{Schur--induced dephasing in the Zeno subspace}
%------------------------------------------------------------

We now turn to the second--order correction generated by the Schur
complement~\eqref{eq:schur}.
Physically, this term describes virtual processes in which a momentum kick
temporarily drives the system away from a fixed $C_Q$--level set, followed by
rapid suppression and return enforced by the monitoring.

As shown in the Appendix~\ref{app:monitoring_zeno}, the net effect of these
processes on the slow Zeno dynamics is an additional pure dephasing of the
momentum--space coherences,
\begin{equation}
\bra{\vec{p}}\Delta\Gen[\rho]\ket{\vec{p}'}
=
-\Gamma_{\rm Schur}(\vec{p},\vec{p}')\,\rho(\vec{p},\vec{p}').
\end{equation}
Crucially, the Zeno elimination singles out the off--diagonal gap
$\Delta C_Q(\vec{p},\vec{p}')$ as the relevant small parameter.
To leading order in $1/\kappa$, the Schur--induced dephasing rate turns out to be linear in
this gap and takes the form
\begin{equation}
\label{eq:schurRateGap}
\Gamma_{\rm Schur}(\vec{p},\vec{p}')
=
\kappa\,\Delta C_Q(\vec{p},\vec{p}')\,
\bigl[C_\Sigma(\vec{p})-C_\Sigma(\vec{p}')\bigr] ,
\end{equation}
where $C_\Sigma(\vec{p})$ is a quadratic form in the momentum.

Related structures appear in general analyses of Zeno--induced effective
generators and constrained open--system dynamics, where the effective
dissipator is linear in the small measurement--induced gap and governed by
second--order Schur complements
\cite{Becker2021}.
In these works, however, the resulting rates are typically expressed in terms
of fixed operators or scalar coefficients determined by the underlying
constraints. 

%------------------------------------------------------------
\subsection{Renormalization of the quadratic observable}
%------------------------------------------------------------

Comparing the Schur--induced dephasing~\eqref{eq:schurRateGap} with the
reference structure~\eqref{eq:deltaLmonRate} obtained from a small shift of
$Q$, we observe that both contributions have exactly the same dependence on
the off--diagonal gap $\Delta C_Q(\vec{p},\vec{p}')$.
This structural equivalence allows a unique identification of the
Schur--complement correction with a renormalization of the monitored
observable,
\begin{equation}
\boxed{
Q\;\longrightarrow\;Q' = Q-\Sigma.
}
\label{eq:coarseGrainedUpdate}
\end{equation}
More precisely, the Schur correction induces a locally defined modification $\Sigma\equiv\Sigma(Q;\pref)$ of the
quadratic form in the vicinity of a reference momentum $\pref$, i.e. $Q'\equiv Q'(Q;\pref)$. In the isotropic Zeno
regime considered below, these local renormalizations are related by symmetry and
may be represented by a single collective quadratic form up to equivalence.

In this sense, the quadratic form $Q$ is not a fixed kinematic input,
but represents a slowly evolving \emph{equivalence class of locally defined
quadratic forms}, whose dynamics is induced by virtual off--shell processes
generated by irreversible momentum--space dynamics.
While measurement--induced renormalization effects have been discussed in
related contexts of monitored open quantum systems
\cite{SarandyLidar2005,Kessler2012,ReiterSorensen2012},
these works primarily address modifications of effective rates, Hamiltonians,
or conditioned state dynamics for a fixed measured observable.
By contrast, the present mechanism leads to a genuine structural
renormalization of the measured quadratic form itself, endowing the Zeno
manifold with an emergent, dynamically generated measurement geometry.

An explicit integral representation of $\Sigma(Q;\pref)$ can be derived by
averaging over the statistics of virtual momentum transfers.
Its detailed form is not essential at this stage and is given for
completeness in Appendix~\ref{app:monitoring_zeno}.

%------------------------------------------------------------
\subsection{Conceptual interpretation}
%------------------------------------------------------------

The central result of this section is that strong continuous monitoring of a
quadratic observable, when combined with irreversible momentum--space
dynamics, induces a systematic backaction on the observable itself.
Virtual off--shell excursions do not merely modify the reduced state within a
fixed Zeno sector, but renormalize the quadratic form that defines the
monitored quantity.

Measurement backaction in continuously monitored open quantum systems has
been extensively analyzed in the contexts of stochastic master equations,
Zeno dynamics, and adiabatic elimination
\cite{wiseman2009quantum,JacobsSteck2006,ChantasriDresselJordan2013,FacchiPascazio2002,AzouitSarletteRouchon2016,SarandyLidar2005}.
In these settings, strong monitoring leads to effective projections,
renormalized Hamiltonians, or reduced Liouvillian generators obtained via
Schur--complement constructions.
Crucially, however, the measured observable is treated as fixed, and
backaction acts at the level of state evolution within a prescribed
measurement geometry.

In contrast, the present construction shows that in the Zeno regime,
the second-order Schur correction generated by irreversible mixing is
structurally equivalent to a variation of the monitored quadratic form,
$Q \rightarrow Q - \Sigma(Q)$.
The effective backaction therefore induces a renormalization flow in the
space of monitored observables rather than a modification of the quantum
state under a fixed operator.

While measurement-induced changes of effective generators are well
established, an endogenous flow of the measured observable itself—generated
by unconditional irreversible dynamics—has not, to our knowledge, been
formulated explicitly in this form.
The mechanism identified here may thus be viewed as a structural extension
of standard Zeno-reduction and adiabatic-elimination techniques, promoting
the monitored observable from a fixed constraint to a dynamical geometric
object.

Iterated over coarse--grained time steps, this mechanism defines an effective
flow in the space of quadratic forms, closely related in spirit to
dynamical coarse-graining approaches in open quantum systems
\cite{schaller2009systematic}.

%============================================================
\section{Renormalization Flow of the Quadratic Observable}\setcounter{equation}{0}
\label{sec:renormalization}
%============================================================

We now turn to the central structural result of this work: the emergence of an
effective renormalization flow for the monitored quadratic observable $Q$.
The previous section established that, in the Zeno regime, virtual off--shell
processes generated by irreversible momentum--space dynamics induce a
well--defined, coarse--grained update~\eqref{eq:coarseGrainedUpdate} of the monitored observable itself.
In this section we interpret this update as the elementary step of a slow
evolution in the space of quadratic forms.

Our aim is to make precise (i) in what sense $Q$ evolves, (ii) how this
evolution follows from the microscopic dynamics via the Zeno separation of
time scales, and (iii) how symmetry and calibration reduce the resulting
dynamics to a small number of effective parameters.

%------------------------------------------------------------
\subsection{Continuum limit and flow equation}
%------------------------------------------------------------

Iterating the update rule~\eqref{eq:coarseGrainedUpdate} over successive
coarse--grained Zeno--elimination steps defines a discrete evolution in the
space of quadratic forms.
Provided the separation of time scales remains intact, such discrete
coarse--grained dynamics admits a continuum limit in which the evolution of $Q$ (understood as a collective representative of the local quadratic form in the Zeno sector) is described by
a differential equation.
This type of discrete--to--continuous limit is standard in repeated
interaction models and continuous limits of stepwise open--system dynamics
\cite{AttalPautrat2006,AttalJoye2005,AzouitSarletteRouchon2016}.

In the continuum limit, the evolution of $Q$ takes the form
\begin{equation}
\label{eq:renorm_flow}
\partial_\lambda Q(\pref) = -\,\mathcal{R}(Q;\pref),
\;
\mathcal{R}(Q;\pref) := \lim_{\delta\lambda\to 0}
\frac{\Sigma(Q;\pref)}{\delta\lambda} .
\end{equation}
Here $\lambda$ denotes a coarse--grained evolution parameter associated with
successive Zeno--elimination steps; it should not be identified with physical
time.

The functional $\mathcal{R}(Q;\pref)$ plays the role of a renormalization rate
for the monitored quadratic observable.
Its explicit form is determined by the statistics of virtual momentum transfers
and by the reference momentum distribution entering the average in the
representation~\eqref{eq:Sigma_representation} for $\Sigma(Q;\pref)$.
Formally, the reduced equation~\eqref{eq:renorm_flow}
defines a flow in the space of quadratic forms.

Importantly, this flow acts on the observable $Q(\pref)$ itself rather than
on the system state $\rho$.
It therefore represents a dynamical evolution in the space of monitored
observables, conceptually distinct from standard dissipative evolution of the
quantum state governed by a fixed Lindblad generator
\cite{wiseman2009quantum,JacobsSteck2006}.

\paragraph*{Remark:}
Measurement backaction is well known to induce systematic modifications
of effective system dynamics in continuously monitored quantum systems,
including measurement-induced heating and diffusion in optomechanical and
cold-atom platforms \cite{ClerkRMP2010,Brahms2012,WeakMeasurementBEC}.
In such settings, the measured operator is treated as fixed, and the
backaction manifests at the level of state evolution, diffusion processes,
or effective Hamiltonian parameters.

More generally, time-dependent effective Lindblad generators arise in the
context of adiabatic open-system dynamics and singular perturbation theory
\cite{SarandyLidar2005,AzouitSarletteRouchon2016,Becker2021},
where slow parameters enter as externally controlled or prescribed
time-dependent quantities.
In these frameworks, the generator evolves, but the monitored observable
remains structurally fixed.

In contrast, the present construction combines strong continuous monitoring
with irreversible momentum mixing in the Zeno regime.
The Schur-induced correction generated by virtual off-shell processes can
be consistently interpreted as a structural renormalization of the monitored
quadratic form itself.
The resulting reduced equation therefore describes an endogenously generated
flow in the space of monitored observables, rather than a dissipative evolution
of the quantum state under a fixed measurement geometry.
In this sense, the dynamics defines a self-induced evolution in the space
of measurement geometries, arising directly from measurement backaction
rather than from externally imposed coarse-graining.

%------------------------------------------------------------
\subsection{Isotropic reduction and parameterization}
%------------------------------------------------------------

The strong monitoring of $C_Q$ suppresses motion in the direction normal to
the level sets $C_Q=\mathrm{const}$.
As a result, Schur--induced corrections can only act within the tangential
subspace.
In isotropic situations, where the momentum--transfer statistics are
rotationally invariant, no preferred directions are singled out within that
subspace.
The induced renormalization therefore reduces to
\begin{equation}
\label{eq:renorm_sigma_iso}
\Sigma(\pref)
=
\sigma_{\rm tan}(\pref)\,\Pi_{\rm tan}(\pref).
\end{equation}
Here $\Pi_{\rm tan}(\pref)$ denotes the local projector onto the tangent space of the
level set $C_Q=\mathrm{const}$ at momentum $\pref$ and $\sigma_{\rm tan}(\pref)$ is a scalar function of $Q$. 

Under the same isotropy assumption, the quadratic form itself can be
parameterized by its normal and tangential components,
\begin{equation}
\label{eq:renorm_Q_param}
Q(\pref)
=
q_{\mathrm{n}}(\pref)\,\Pi_{\mathrm{n}}(\pref)
+
q_{\mathrm{tan}}(\pref)\,\Pi_{\mathrm{tan}}(\pref),
\end{equation}
where $\Pi_{\mathrm{n}}(\pref)$ projects onto the normal direction singled out by
$\nabla|_{\pref} C_Q$ and $\Pi_{\mathrm{tan}}(\pref)$ onto its orthogonal complement,
$\Pi_{\mathrm{n}}(\pref)+\Pi_{\mathrm{tan}}(\pref)=\mathbbm 1$.
This decomposition makes explicit that the renormalization flow affects only
the relative weighting of normal and tangential directions.

%------------------------------------------------------------
\subsection{Calibration and reduction of degrees of freedom}
%------------------------------------------------------------

A crucial point is that the overall scale of the quadratic form is not a physically
observable quantity. After restriction to the Zeno sector, the effective object is a
family of locally defined quadratic forms $Q(\pref)$, related by symmetry transport and
representing a single slow collective degree of freedom. A local rescaling
$Q(\pref)\mapsto \alpha\,Q(\pref)$ merely rescales the monitored quantity
$C_Q(\pref,\deltap)$ and can be absorbed into a corresponding change of the effective
monitoring strength. The renormalization flow~\eqref{eq:renorm_flow} must therefore be
considered modulo such rescalings, which act uniformly on all representatives of the
equivalence class.

To fix this redundancy, we introduce a calibration condition that selects a representative
of the equivalence class by fixing the characteristic measurement scale generated by the
Zeno dynamics itself.

\subsection{Local measurement contrast and emergent geometry}

After restriction to the Zeno sector, the relevant information carried by the monitored
observable is not the absolute value of $C_Q(p)$, but its local variation around a
reference momentum $\pref$. For near--diagonal coherences,
$\vec{p}_{1,2}=\pref\pm\tfrac12\deltap$ with $\|\deltap\|\ll 1$, we define the local
measurement contrast
\begin{equation}
\label{eq:local_epsilon}
\begin{split}
\varepsilon(\pref,\deltap)
&:= C_{Q}\!\left(\pref+\tfrac12\deltap\right)
 - C_{Q}\!\left(\pref-\tfrac12\deltap\right)\\
&= 2\bigl(Q(\pref)\,\pref\bigr)\cdot\deltap
 = \nabla C_{Q}(\pref)\cdot\deltap .
\end{split}
\end{equation}
This exact identity follows from the quadratic form of $C_Q$ and induces a natural local
decomposition of momentum increments into a normal direction
$\vec{n}(\pref)=2Q(\pref)\,\pref$ and a tangential subspace defined by
$\vec{n}(\pref)\cdot\deltap=0$.

\subsection{Zeno damping scale and calibration}

The monitoring generator suppresses coherences between momentum eigenstates
$\ket{\pref+\tfrac12\deltap}$ and $\ket{\pref-\tfrac12\deltap}$ at a rate
$(\kappa/2)\,\varepsilon(\pref,\deltap)^2$. This is seen directly from the matrix element
$\bra{\vec{p}_1}\Gen{mon}[\rho]\ket{\vec{p}_2}$ and controls the decay of off--diagonal
modes as well as the denominator of the Zeno resolvent entering the Schur complement.

Accordingly, the natural local scale generated by the monitoring dynamics is the
characteristic local Zeno damping rate
\begin{equation}
\label{eq:local_Gamma}
\Gamma(\pref,\deltap)
:= \frac{\kappa}{2}\,\varepsilon(\pref,\deltap)^2 ,
\end{equation}
evaluated on the Zeno scale $\|\deltap\|\sim O(\kappa^{-1/2})$. Importantly, $\Gamma(\pref)$
depends on $Q(\pref)$ only through its overall scale and carries no additional geometric
information. Fixing $\Gamma(\pref)$ therefore in principle fixes the rescaling freedom of the local
quadratic form without constraining its tangential structure. To support it with physical meaning this quantity has to be averaged over all relevant realization of $\deltap$.

\subsection{Choice of representative and reduction of degrees of freedom}

Formally, the renormalization flow is defined on the projective space 
of nondegenerate quadratic forms modulo positive rescaling.
The calibration procedure selects a representative within each projective class.

In the homogeneous and isotropic regime considered below, the local quadratic forms
$Q(\pref)$ are equivalent up to symmetry transport, and the Zeno damping scale
$\Gamma(\pref)$ is independent of the specific representative $\pref$. Calibration is
implemented by rescaling $Q(\pref)$ after each coarse--grained update of the flow such
that the characteristic Zeno scale~\eqref{eq:local_Gamma} is kept fixed. This procedure
merely fixes the overall unit of the reduced dynamics with respect to the evolution parameter
$\lambda$ and does not affect the structure of the renormalization flow or its fixed points.

With this choice of representative, the effective dynamics reduces to a single
dimensionless parameter characterizing the equivalence class of local quadratic forms,
namely the ratio of tangential and normal weights,
\begin{equation}
\label{eq:renorm_ratio}
r := \frac{q_{\mathrm{tan}}}{q_{\mathrm{n}}},
\end{equation}
where $q_{\mathrm{n}}$ and $q_{\mathrm{tan}}$ denote the locally defined normal and
tangential coefficients of $Q(\pref)$. The reduced flow for $r$ is invariant under local
rescalings of the representative and captures the intrinsic infrared behavior of the
Zeno--induced renormalization.

Physically, the calibration expresses a self--consistency requirement: in a stationary
monitoring environment that is not modified by the system, the characteristic measurement
scale generated by the monitoring process must remain fixed under the effective dynamics.
Operationally, this ensures that the Zeno gap and the associated separation of time scales
are preserved, in close analogy to normalization procedures in continuous quantum
measurement and adiabatic elimination
\cite{wiseman2009quantum,SarandyLidar2005,AzouitSarletteRouchon2016}.

%============================================================
\section{Fixed point structure and emergent Lorentzian signature}
\label{sec:fixedpoint}
%============================================================

We now analyze the effective renormalization dynamics induced by the
coarse--grained Zeno update together with a \emph{quadratic} local calibration
prescription. The key point is that, after restriction to the Zeno sector, the
monitored quadratic form is defined only up to local rescalings
$Q(\pref)\mapsto \alpha\,Q(\pref)$, and the dynamics admits a natural reduction
to a one--dimensional projective flow.

Importantly, the overall scale  is not a physical quantity. Indeed,
a local rescaling  leaves the structure of
the Zeno sector invariant and merely rescales the measurement
contrast $\varepsilon(\pref,\deltap)$ and thus the associated decay rate. What is
physically meaningful is not the normalization of $Q(\pref)$  itself,
but the characteristic time scale generated by the monitoring
environment. Calibration therefore fixes the local rescaling freedom by
requiring that the monitoring–induced Zeno damping rate re
mains invariant under the effective flow. 

\subsection*{Local Zeno averaging and quadratic calibration scale}

For fixed $\pref$, the microscopic monitoring contrast is given by~\eqref{eq:local_epsilon}
while~\eqref{eq:local_Gamma} defines the characteristic local Zeno damping rate.
The \emph{physically relevant} local Zeno damping scale is obtained by averaging over
the \emph{relevant} (near--diagonal) Zeno increments $\deltap$.
We denote this averaging by $\langle\cdot\rangle_{\rm rel}$ and set
\begin{equation}
\label{eq:def_C_quadratic}
\Gamma(\pref)
:=
\big\langle \Gamma(\pref,\deltap)\big\rangle_{\rm rel}
=
\int_{\mathbb R^4} \frac{\kappa}{2}\,\varepsilon(\pref,\deltap)^2\,
\mu_{\pref}(d\deltap),
\end{equation}
where $\mu_{\pref}$ is a normalized local increment law (see Appendix~\ref{app:localProjectors} for technical details).
While the Zeno monitoring constrains fluctuations only in the normal
direction, tangential momentum increments are statistically bounded by the
underlying mixing dynamics. In the framework of the quantum linear Boltzmann
equation, this corresponds to the standard assumption of finite second moments
of the momentum transfer distribution, which is required for the generator to
be well defined. The effective averaging therefore remains local and finite
without imposing an explicit cutoff.

Since $\varepsilon(\pref,\deltap)$  depends linearly on the quadratic form, the corresponding decay rate scales quadratically under local rescalings $Q(\pref)\mapsto \alpha_{\rm cal}\,Q(\pref)$:
\begin{equation}
\label{eq:C_scales_quadratic}
\Gamma(\pref)\mapsto\alpha_{\rm cal}^2\,\Gamma(\pref).
\end{equation}
We will use this invariance of the Zeno rate $\Gamma(\pref)$ to fix the local rescaling freedom.

\subsection*{Local update and calibration}

On one coarse--grained step $\delta\lambda$, the raw Zeno update of the monitored
quadratic form is
\begin{equation}
\label{eq:raw_update_sec5}
Q(\pref)\;\longmapsto\;\widetilde Q(\pref)=Q(\pref)-\Sigma \bigl(\pref\bigr),
\end{equation}
where $\Sigma(\pref)$ is the Schur--complement correction derived in Sec.~3 and
acts only in the tangential subspace at $\pref$.

Calibration compensates the raw update by a subsequent rescaling
\begin{equation}
Q^+(\pref):=\alpha_{\rm cal}(\pref)\,\widetilde Q(\pref),
\label{eq:calibrationOfQ}
\end{equation}
where the positive factor $\alpha_{\rm cal}(\pref)$ is chosen such that the
\emph{quadratic} damping scale \eqref{eq:def_C_quadratic} is preserved,
\begin{equation}
\label{eq:calibration_condition}
\Gamma^+(\pref)=\Gamma(\pref).
\end{equation}
Since $\Gamma$ scales quadratically under $Q\mapsto \alpha Q$, this yields
\begin{equation}
\label{eq:alpha_cal_quadratic_exact}
\alpha_{\rm cal}(\pref)=\sqrt{\frac{\Gamma(\pref)}{\widetilde \Gamma(\pref)}}.
\end{equation}
To express $\widetilde \Gamma$ to first order in the tangential Schur increment, we
use the local parametrization~\eqref{eq:renorm_Q_param} of $Q$ and by the isotropy assumption~\eqref{eq:renorm_sigma_iso} get
\begin{equation}
q_{\mathrm n}\mapsto \widetilde q_{\mathrm n}=q_{\mathrm n},
\qquad
q_{\mathrm{tan}}\mapsto \widetilde q_{\mathrm{tan}}=q_{\mathrm{tan}}-\sigma_{\mathrm{tan}}.
\label{eq:updateRuleCoefficientsQ}
\end{equation}
The averaged quantity $\Gamma(\pref)$ depends on $Q(\pref)$ only through the local
coefficients $q_{\mathrm n}(\pref)$ and $q_{\mathrm{tan}}(\pref)$.
Moreover, by \eqref{eq:local_epsilon}--\eqref{eq:local_Gamma} and the
definition \eqref{eq:def_C_quadratic}, $\Gamma(\pref)$ is homogeneous of degree two in
$(q_{\mathrm n},q_{\mathrm{tan}})$. We therefore introduce the local scalars
\begin{equation}
\label{eq:defAandB_quadratic}
A(\pref):=\frac{\partial \Gamma(\pref)}{\partial q_{\mathrm n}(\pref)},
\qquad
B(\pref):=\frac{\partial \Gamma(\pref)}{\partial q_{\mathrm{tan}}(\pref)}.
\end{equation}
By Euler's theorem for homogeneous functions of degree two, this implies the
\emph{exact} identity
\begin{equation}
\label{eq:defineC_quadratic_Euler}
\Gamma(\pref)
=\frac{1}{2}\Bigl(
q_{\mathrm n}(\pref)\,A(\pref)
+
q_{\mathrm{tan}}(\pref)\,B(\pref)\Bigr).
\end{equation}
In terms of the local increment moments with respect to $\mu_{\pref}$, one may
write this more explicitly. Denoting the normal and tangential components $(\mathrm{n},\mathrm{tan})$ by $(x,y)$ and setting $p_{\mathrm x/y}:=\Pi_{\mathrm x/y}(\pref)\,\pref$, we define the second moments
\begin{equation}
\label{eq:M_moments}
M_{\mathrm{xy}}(\pref):=\big\langle (p_{\mathrm x}\!\cdot\deltap)(p_{\mathrm{y}}\!\cdot\deltap)\big\rangle_{\rm rel}.
\end{equation}
Then employing the definition for the local measurement contrast~\eqref{eq:local_epsilon} and the characteristic local Zeno damping rate~\eqref{eq:local_Gamma} one finds
\begin{equation}
\label{eq:C_explicit_quadratic}
\begin{split}
\Gamma(\pref)
&=
2\kappa\,\Big(
q_{\mathrm n}(\pref)^2\,M_{nn}(\pref)
+q_{\mathrm{tan}}(\pref)^2\,M_{tt}(\pref)\\
&\qquad\qquad +2q_{\mathrm n}(\pref)q_{\mathrm{tan}}(\pref)\,M_{nt}(\pref)
\Big),
\end{split}
\end{equation}
and therefore~\eqref{eq:defAandB_quadratic} yields
\begin{equation}
\begin{split}
\label{eq:A_B_explicit_quadratic}
A(\pref)&=4\kappa\Big(q_{\mathrm n}M_{nn}+q_{\mathrm{tan}}M_{nt}\Big),\\
B(\pref)&=4\kappa\Big(q_{\mathrm{tan}}M_{tt}+q_{\mathrm n}M_{nt}\Big).
\end{split}
\end{equation}

Finally, by~\eqref{eq:updateRuleCoefficientsQ} into~\eqref{eq:C_explicit_quadratic} the first--order variation of $\Gamma$ under the raw tangential shift
$\delta q_{\mathrm{tan}}=-\sigma_{\mathrm{tan}}$ is
\begin{equation}
\label{eq:C_tilde_firstorder}
\widetilde \Gamma(\pref)
=
\Gamma(\pref)-\sigma_{\mathrm{tan}}(\pref)\,B(\pref)
+\mathcal O\bigl(\sigma_{\mathrm{tan}}^2\bigr),
\end{equation}
so that \eqref{eq:alpha_cal_quadratic_exact} becomes, to first order,
\begin{equation}
\label{eq:alpha_cal_quadratic_firstorder}
\begin{split}
\alpha_{\rm cal}(\pref)
&=
\sqrt{\frac{\Gamma(\pref)}{\Gamma(\pref)-\sigma_{\mathrm{tan}}(\pref)\,B(\pref)}}\\
&=
1+\frac{\sigma_{\mathrm{tan}}(\pref)}{2}\frac{B(\pref)}{\Gamma(\pref)}
+\mathcal O\bigl(\sigma_{\mathrm{tan}}^2\bigr).
\end{split}
\end{equation}

All relevant contributions so far are built from positive objects such as rates, expectation values and positive numbers. 
The calibration~\eqref{eq:calibrationOfQ} thus rescales both coefficients of $Q$ by
the same positive factor $\alpha_\mathrm{cal}$ and particularly does not affect the dimensionless ratio
\begin{equation}
\label{eq:r_def_sec5}
r(\pref):=\frac{q_{\mathrm{tan}}(\pref)}{q_{\mathrm n}(\pref)}.
\end{equation}
The dynamically relevant evolution therefore reduces to a one--dimensional flow for
$r$ following from the raw update
\begin{equation}
\label{eq:rawUpdate_r}
r\mapsto r-\sigma_{\mathrm{tan}}/q_{\mathrm n}.
\end{equation}

\subsection*{Continuum limit and fixed point}
Passing to the continuum limit $\delta\lambda\to0$, we assume that the local
Schur increment scales linearly with the coarse--graining step,
\begin{equation}
\label{eq:renormalizationRate}
\sigma_{\mathrm{tan}}(r;\pref)
=
\rho_{\mathrm{tan}}(r;\pref)\,\delta\lambda
+
o(\delta\lambda),
\end{equation}
where, due to the positivity of $\sigma_{\mathrm{tan}}$, the renormalization rate $\rho_{\mathrm{tan}}(r;\pref)$ must be a positive function of $r$ and $\pref$. Using \eqref{eq:defineC_quadratic_Euler} and $q_{\mathrm{tan}}=rq_{\mathrm n}$ one has
\begin{equation}
\label{eq:qn_from_C}
q_{\mathrm n}(\pref)=\frac{2\,\Gamma(\pref)}{A(\pref)+r(\pref)\,B(\pref)},
\end{equation}
and, hence the continuum flow takes the form
\begin{equation}
\label{eq:r_flow_quadratic}
\boxed{
\partial_\lambda r
=
-\rho_{\mathrm{tan}}(r;\pref)\,
\frac{A(\pref)+r\,B(\pref)}{2\,\Gamma(\pref)} .
}
\end{equation}
The stationary solution of \eqref{eq:r_flow_quadratic}, at which the flow
vanishes, is
\begin{equation}
\label{eq:r_fixed_quadratic}
\boxed{
r_\star=-\frac{A}{B}.
}
\end{equation}
Whenever $A(\pref),B(\pref)>0$, the fixed point lies at a strictly negative
value of $r$.

\subsection*{Emergent Lorentzian signature and stability}

At the fixed point, the calibrated quadratic form can be written as
\begin{equation}
\label{eq:Q_star_quadratic}
Q_\star(\pref)
=
q_{\mathrm n}(\pref)\Bigl(\Pi_{\mathrm n}(\pref)+r_\star\,\Pi_{\mathrm{tan}}(\pref)\Bigr),
\end{equation}
which has one eigenvalue of opposite sign relative to the remaining three. In an
adapted local basis it takes the diagonal form
\[
\boxed{
Q_\star
\sim
\mathrm{diag}\!\left(1,-\alpha_{\mathrm L},-\alpha_{\mathrm L},-\alpha_{\mathrm L}\right),
\qquad
\alpha_{\mathrm L}:=-r_\star>0,
}
\]
up to an overall scale fixed by calibration.

Linearizing \eqref{eq:r_flow_quadratic} around $r_\star$ yields
\[
\partial_\lambda(r-r_\star)
=
-\rho_{\mathrm{tan}}(r_\star)\,
\frac{B}{2\,\Gamma}\,(r-r_\star)
+
\mathcal O\bigl((r-r_\star)^2\bigr),
\]
so that the fixed point is attractive whenever $\rho_{\mathrm{tan}}(r_\star)>0$.
The nonvanishing linear term implies that $r_\star$ is hyperbolic and therefore
structurally stable under small perturbations of the reduced dynamics.

\paragraph*{Quantitative structure of the fixed point}

Under the isotropy assumptions underlying the local Schur reduction and the
standard framework of the quantum linear Boltzmann equation, the mixed normal--
tangential moment vanishes identically,
$M_{nt}(\pref)=0$, and the tangential fluctuations are isotropic within the
three--dimensional Zeno tangent space.
In this case the fixed--point condition \eqref{eq:r_fixed_quadratic} reduces to
the self--consistency relation
\begin{equation}
\label{eq:rstar_moments}
r_\star^2
=
\frac{M_{nn}(\pref)}{M_{tt}(\pref)},
\end{equation}
where $M_{nn}$ and $M_{tt}$ denote the locally averaged second moments of the
relevant momentum increments along the normal and tangential directions,
respectively.

Within the standard assumptions of the quantum linear Boltzmann equation---in
particular finite temperature, finite scattering cross sections, and the
existence of finite second moments of the momentum transfer distribution---both
moments are strictly positive and finite.
As a consequence, the fixed point $r_\star<0$ is guaranteed to exist and to be
finite, with $|r_\star|$ generically of order unity.
While its precise numerical value depends on microscopic scattering details,
in close analogy to transport coefficients such as diffusion constants, its
existence and Lorentzian character are robust and do not rely on fine tuning or
additional symmetry assumptions.

Importantly, no relativistic symmetry is assumed at the microscopic level. The
underlying irreversible dynamics is formulated in a purely Euclidean
four--dimensional momentum space. The emergence of a Lorentzian signature is an
infrared property of the locally calibrated, projective renormalization flow,
resulting solely from the interplay of continuous monitoring, dissipation, and
calibration, rather than from fine tuning or imposed kinematic symmetry.

%============================================================
\section{Physical Interpretation and Stationary States}\setcounter{equation}{0}
\label{sec:interpretation}
%============================================================

The previous section established the existence of a stable infrared fixed point of the calibrated Zeno-induced renormalization flow for the monitored quadratic form. Under isotropy and the quadratic calibration prescription, the effective dynamics reduces to a one-dimensional projective flow whose unique attractive fixed point is characterized by a Lorentzian signature. Importantly, this result was obtained without assuming any relativistic symmetry at the microscopic level; it arises solely from the interplay of irreversible momentum-space mixing, strong continuous monitoring, and calibration.

Having identified the fixed-point structure at the level of the monitored observable, we now turn to its physical interpretation. The goal of this section is twofold. First, we reinterpret the Lorentz-signature quadratic form emerging at the infrared fixed point in geometric terms and clarify how Euclidean equivalence classes are deformed into hyperbolic ones under the renormalization flow. Second, we analyze the consequences of this fixed-point structure for the Zeno-projected long-time dynamics, including the emergence of mass-shell–like constraints and the form of stationary momentum-space distributions under detailed-balance conditions.

\subsection{From Euclidean equivalence to emergent Lorentz kinematics}
\label{subsec:euclid_to_lorentz}

At $\lambda=0$ the underlying momentum space is purely Euclidean.
The quadratic invariant is simply
\[
\vec{p}^2 = \vec{p}^{\mathsf T} \vec{p} ,
\]
and all momentum vectors with equal norm lie on the same
three--sphere in $\mathbb R^4$.
The symmetry group relating such momenta is the Euclidean rotation group
$O(4)$.
Hence, at $\lambda=0$, systems with equal $p^2$ are trivially equivalent:
they differ only by a global $O(4)$ rotation.

\paragraph{Isotropic flow and preservation of equivalence}

We assume that the microscopic dynamics (mixing plus monitoring) is
translation invariant and isotropic in $\mathbb R^4$.
Under these conditions, both the renormalization flow and the calibration
procedure depend only on scalar contractions and therefore act identically
on all momenta with the same initial norm.
Consequently, if two systems satisfy
\[
\vec{p}_1^2 = \vec{p}_2^2 \quad \text{at } \lambda=0 ,
\]
then at any finite $\lambda$ one has
\[
\vec{p}_1^{\mathsf T} Q(\lambda) \vec{p}_1
=
\vec{p}_2^{\mathsf T} Q(\lambda) \vec{p}_2
= m^2(\lambda) .
\]
Thus the flow preserves the equivalence class defined by the initial
Euclidean invariant.

\paragraph{Emergence of hyperbolic geometry}

As $\lambda$ increases, the quadratic form $Q(\lambda)$ evolves
towards the Lorentzian fixed point $Q_\star$.
The invariant surfaces
\[
\vec{p}^{\mathsf T} Q_\star \vec{p} = m^2
\]
are no longer three--spheres but hyperboloids.
The original Euclidean equivalence under $O(4)$ is replaced by
invariance under the isometry group of $Q_\star$,
\[
\Lambda^{\mathsf T} Q_\star \Lambda = Q_\star ,
\]
which is isomorphic to $O(1,3)$ after normalization.

In this sense the renormalization flow deforms the geometry of
momentum space from spherical to hyperbolic,
without introducing any preferred direction.
The equivalence class of momenta with fixed $m^2$ survives,
but its symmetry group changes.

\paragraph{Choice of inertial system}

Selecting a representative momentum
$\vec{u} \in \mathcal M_m := \{\vec{p} \mid \vec{p}^{\mathsf T} Q_\star \vec{p} = m^2\}$
corresponds to choosing an inertial system.
There exists a Lorentz transformation $\Lambda_\vec{u}$
such that
\[
\Lambda_\vec{u} \vec{u} = (m,0,0,0) .
\]
Relative to this choice, any other momentum
$p \in \mathcal M_m$
is described by
\[
[\vec{p}]_\vec{u} := \Lambda_\vec{u} \vec{p}.
\]
Different choices of representative correspond to different inertial
frames and are related by Lorentz transformations.

\paragraph{Physical picture}

Initially, equivalence of systems is governed by Euclidean rotations in the
underlying momentum space.
Under the coarse--grained renormalization flow, the quadratic invariant
characterizing the monitored dynamics is continuously deformed.
At the infrared fixed point, this invariant acquires Lorentzian signature,
and its level sets become relativistic mass shells.

In this sense, the original $O(4)$ equivalence of the Euclidean background
is replaced, at the level of the renormalized quadratic form, by the
isometry group of $Q_\star$, which is locally isomorphic to $O(1,3)$.
Lorentz transformations thus emerge not as fundamental symmetries of the
microscopic theory, but as the natural isometries of a representative of
the infrared projective equivalence class selected by the Zeno flow.

In this way, relativistic kinematics appears as an infrared geometric
property of an underlying Euclidean momentum space.

\subsection{Mass--shell constraints and Zeno--projected dynamics}

At the infrared fixed point, the quadratic observable
\[
C_{Q_\star}(\opvec{p})=\opvec{p}^{\mathrm T}Q_\star \opvec{p}
\]
defines a hyperbolic level--set structure in the extended momentum space.
Within the present framework, these level sets function as effective
mass--shell--like constraint surfaces.
Importantly, they enter only at the level of the Zeno--projected description
and are not imposed as kinematic constraints on the microscopic dynamics.

In the strong--monitoring regime, motion normal to the level sets is strongly
suppressed by the quantum Zeno effect
\cite{misra1977zeno,FacchiPascazio2002,Becker2021},
while tangential motion along the level sets remains dynamically active.
As a consequence, the effective long--time dynamics generated by the reduced
evolution is confined to the vicinity of a fixed level set of
$C_{Q_\star}(p)$.
In this sense, mass--shell constraints appear here as effective restrictions
on the accessible momentum--space dynamics induced by monitoring and
irreversible coupling, rather than as fundamental postulates.

\subsection{Energy scale, effective evolution and physical time}

The flow parameter $\lambda$ governing the renormalization of the quadratic form 
parametrizes successive coarse–graining and elimination steps. It measures how far fast off–shell 
modes have been integrated out and thus plays the role of a renormalization 
scale rather than that of a dynamical evolution parameter.

At the infrared fixed point, the quadratic form $Q_\star$ acquires Lorentzian 
signature. This structural property is a mathematical consequence of the 
calibrated Zeno-induced flow and does not rely on any a priori relativistic 
assumption.

Whether one may interpret the distinguished \emph{timelike} direction of $Q_\star$ as an 
emergent energy variable and introduces a corresponding notion of physical 
time is an additional interpretative step. The present analysis does not 
require such an identification; it merely establishes that the effective 
infrared geometry admits a Lorentzian structure compatible with standard 
relativistic kinematics. 

In this sense, the appearance of Lorentzian signature is a dynamical result, 
while its interpretation in terms of physical time and energy is a question of interpretation.

\subsection{Equilibrium distributions}

We finally discuss the structure of stationary momentum--space distributions
associated with the effective dynamics in the vicinity of the Lorentz--signature
fixed point.
Throughout this subsection, we work within the restricted scope of the
Zeno--projected, coarse--grained dynamics and do not address questions of
fundamental equilibration or spacetime symmetry.

The effective evolution in momentum space is generated by a translation--covariant
GKSL operator whose dissipative part arises from irreversible momentum--transfer
processes.
Under standard detailed--balance conditions for these processes, the corresponding
generator admits stationary states of Gibbs type
\cite{davies1974markovian,BreuerPetruccioneBook,Vacchini2001}.
Concretely, detailed balance is understood here in the sense that the momentum--transfer
rates satisfy a local Kubo--Martin--Schwinger (KMS) -- type condition with respect to a distinguished energy
functional, ensuring the absence of stationary probability currents in momentum space.

\paragraph{Additional assumption (thermal rest frame / bath four--velocity)}
To identify a Maxwell--J\"uttner equilibrium, one needs, in addition to the Lorentz--signature
fixed point, a \emph{frame selection} specifying the energy with respect to which the environment
thermalizes the momentum degrees of freedom. We encode this by assuming that the effective
rates satisfy detailed balance with respect to a bath four--velocity $u$ (the rest frame of the
environment), i.e.\ with respect to the scalar energy functional
\begin{equation}
\label{eq:energy_u_def}
E_u(p) := u\cdot p,
\end{equation}
where $\cdot$ denotes the bilinear form induced by $Q_\star$ (in an adapted frame, $E_u(p)=p^0$).
This assumption is standard in relativistic kinetic theory and represents the
Markovian/KMS input selecting the equilibrium notion; it is \emph{not} implied by the fixed point
mechanism itself.

\paragraph{Derivation of the Gibbs form from detailed balance}
To make the appearance of Maxwell--J\"uttner--type equilibria explicit, we focus on the
diagonal sector of the reduced dynamics, i.e.\ on momentum populations
$f(p):=\langle \vec{p}|\rho|\vec{p}\rangle$. For translation--covariant momentum--transfer generators
of Boltzmann type (as obtained from the QLBE on the diagonal), the populations satisfy a
classical master equation
\begin{equation}
\label{eq:ME_pop}
\partial_\lambda f(\vec{p})
=
\int d^4\kick\;
\Bigl[
W(\vec{p}-\kick,\kick)\,f(\vec{p}-\kick)
-
W(\vec{p},\kick)\,f(\vec{p})
\Bigr],
\end{equation}
where $W(\vec{p},\kick)\ge 0$ denotes the transition rate density for a momentum jump
$\vec{p}\mapsto \vec{p}+\kick$.

We assume \emph{detailed balance} with respect to the bath energy $E_u$, in the standard sense
that the forward and backward rates satisfy
\begin{equation}
\label{eq:DB_rates_u}
W(\vec{p},\kick)\,e^{-\beta E_u(\vec{p})}
=
W(\vec{p}+\kick,-\kick)\,e^{-\beta E_u(\vec{p}+\kick)},
\end{equation}
for all $\vec{p},\kick$, and for some $\beta>0$.
This condition expresses the absence of stationary probability currents in momentum space and
is the natural Markovian formulation of thermal equilibrium for jump processes.

Define
\begin{equation}
\label{eq:Gibbs_candidate_u}
f_\beta(\vec{p}):=Z^{-1}e^{-\beta E_u(\vec{p})}.
\end{equation}
Then, inserting $f_\beta$ into \eqref{eq:ME_pop} and using \eqref{eq:DB_rates_u} yields
\[
W(\vec{p}-\kick,\kick)\,f_\beta(\vec{p}-\kick)
=
W(\vec{p},-\kick)\,f_\beta(\vec{p}),
\]
hence the gain and loss terms cancel pairwise under the $\kick$--integration, and one obtains
$\partial_\lambda f_\beta(\vec{p})=0$ for all $\vec{p}$. Therefore, $f_\beta$ is a stationary solution
of the population dynamics. Under standard irreducibility assumptions on the jump process
(connectivity in momentum space), detailed balance furthermore implies uniqueness of the stationary
distribution within the normalization class.

\paragraph{Identification with Maxwell--J\"uttner at the Lorentz fixed point}
At the infrared fixed point of the calibrated Zeno flow, the effective quadratic form $Q_\star$
has Lorentzian signature. Its level sets define a distinguished mass shell
\begin{equation}
\label{eq:mass_shell_Qstar}
\mathcal M_m := \bigl\{\,p\in\mathbb R^4 \;\big|\; p^{\trans}Q_\star p = m^2,\; p^0>0 \,\bigr\},
\end{equation}
and Zeno suppression confines the long--time dynamics to (a neighborhood of) such a shell.

In an adapted local frame in which $Q_\star\sim \mathrm{diag}(1,-\alpha,-\alpha,-\alpha)$ and for a bath
at rest, $u=(1,0,0,0)$, one has
\begin{equation}
\label{eq:Eu_restframe}
E_u(p)=p^0,
\qquad
\text{and on }\mathcal M_m:\quad
p^0=\sqrt{m^2+\alpha\,\mathbf p^2}.
\end{equation}
With this choice, the Gibbs stationary state \eqref{eq:Gibbs_candidate_u} becomes
\begin{equation}
\label{eq:MJ_from_p0}
f_\beta(p)\propto \exp\bigl(-\beta\,p^0\bigr),
\end{equation}
which is the Maxwell--J\"uttner form (up to the conventional choice of reference measure on the mass shell,
e.g.\ whether $f$ is taken as a density with respect to $d^3\mathbf p$ or the invariant mass--shell measure
$d^3\mathbf p/p^0$).

\paragraph{Remark (why $\sqrt{p^{\trans}Q_\star p}$ is not the energy)}
Note that the Lorentz--invariant scalar $\sqrt{p^{\trans}Q_\star p}$ equals $m$ on $\mathcal M_m$ and is thus
constant along the Zeno--constrained mass shell. Therefore, it cannot serve as the thermalizing energy functional
for a nontrivial equilibrium distribution on $\mathcal M_m$. The appropriate energy entering detailed balance is the
frame--dependent quantity $E_u(p)=u\cdot p$, selected by the environment.

This class of distributions is well known from relativistic kinetic theory and from relativistic Boltzmann equations
\cite{deGroot1980relativistic,dunkel2007relativistic}.
Their appearance here does not rely on an explicit relativistic invariance of the microscopic dynamics, but follows from:
(i) the emergence of a Lorentz--signature quadratic form $Q_\star$ selecting a mass--shell geometry, and
(ii) an additional detailed--balance (KMS) assumption with respect to a bath rest frame $u$ that picks out the energy
component $E_u$.

In the nonrelativistic regime (small spatial momenta in the adapted frame), \eqref{eq:Eu_restframe} reduces to
$p^0 \approx m + \frac{\alpha}{2m}\mathbf p^2$, so that \eqref{eq:MJ_from_p0} approaches the usual Gaussian equilibrium
(up to the irrelevant constant $e^{-\beta m}$), consistent with the standard QLBE stationary states
\cite{VacchiniHornberger2009}, providing a consistency check with established open--system descriptions of irreversible
momentum--space dynamics.

Taken together, these results show that the fixed--point structure of the calibrated Zeno dynamics is compatible with
both relativistic and nonrelativistic equilibrium theory, while remaining firmly within the scope of an effective
open--system description.

\subsection*{Summary and outlook}

Within the extended quantum linear Boltzmann framework considered here, the
interplay of irreversible momentum--space dynamics, strong continuous
monitoring, and calibration leads to a stable infrared fixed point of the
monitored quadratic observable.
This fixed point is characterized by a Lorentzian signature and induces
mass--shell--like constraints at the level of the Zeno--projected effective
dynamics.

The present analysis is restricted to a specific class of Markovian open
systems and monitoring schemes and does not constitute a fundamental
derivation of relativistic symmetry.
Possible extensions include anisotropic monitoring, interacting systems, and
dynamical environments.
More generally, the results illustrate how kinematic constraints can arise as
effective features of monitored open--system dynamics within a concrete and
experimentally motivated model class.

%=====================================================================
\section{Appendix: Local Schur Correction and Identification of the Renormalized Quadratic Form
\texorpdfstring{$Q\mapsto Q-\Sigma(Q)$}{Q -> Q - Sigma(Q)}}\setcounter{equation}{0}
\label{app:monitoring_zeno}
%=====================================================================

This section derives, within a translation-covariant quantum linear Boltzmann framework,
the leading Zeno-limit Schur correction induced by strong continuous monitoring of a
quadratic momentum observable. The result is a closed and explicit identification of the
second-order correction with a renormalization of the monitored quadratic form, $Q \longmapsto Q-\Sigma(Q)$
together with a representation of $\Sigma(Q)$ in terms of the microscopic QLBE kernel.
Throughout, $\eta$ denotes the weak coupling parameter to the environment generator, while
$\kappa\gg 1$ is the monitoring strength.

\subsection{Monitored observable and Zeno band}
\label{subsec:zeno_band}

Let $\opvec{P}\in\mathbb{R}^4$ and $\opvec{X}\in\mathbb{R}^4$ denote the momentum and position operators of a test particle
(with the usual CCR), and let $C_Q(\opvec{P}):=\opvec{P}^\top Q \opvec{P}$ be the monitored quadratic observable parametrized by the quadratic form $Q=Q^\top\in\mathbb{R}^{4\times 4}$. Continuous monitoring is modelled by the pure-dephasing GKSL generator
\begin{equation}
\mathcal{L}^{Q}_{\mathrm{mon}}[\rho]
:=-\frac{\kappa}{2}\,[C_Q,[C_Q,\rho]],
\qquad \kappa\gg 1.
\label{eq:Lmon_def}
\end{equation}
In the momentum representation $\rho(\vec{p}_1,\vec{p}_2)=\langle \vec{p}_1|\rho|\vec{p}_2\rangle$, $C_Q$ is diagonal and one finds
\begin{equation}
\bigl(\mathcal{L}^{Q}_{\mathrm{mon}}\rho\bigr)(\vec{p}_1,\vec{p}_2)
=-\frac{\kappa}{2}\,\bigl(\Delta C_Q(\vec{p}_1,\vec{p}_2)\bigr)^2\,\rho(\vec{p}_1,\vec{p}_2),
\label{eq:Lmon_mom}
\end{equation}
where $\Delta C_Q(\vec{p}_1,\vec{p}_2):=C_Q(\vec{p}_1)-C_Q(\vec{p}_2)$ defines the gap between two different measurement outcomes at $\vec{p}_1$ and $\vec{p}_2$. Hence, in momentum representation the semigroup generated by $\mathcal{L}^{Q}_{\mathrm{mon}}$ acts as a Gaussian damping factor,
\begin{equation}
\bigl(e^{t\mathcal{L}^{Q}_{\mathrm{mon}}}\rho\bigr)(\vec{p}_1,\vec{p}_2)
=e^{-\frac{\kappa t}{2}\bigl(\Delta C_Q(\vec{p}_1,\vec{p}_2)\bigr)^2}\rho(\vec{p}_1,\vec{p}_2),
\label{eq:semigroup_mon}
\end{equation}
implicating that, on the slow (Zeno) time scale, coherences survive only if their monitoring gap is small. This suppression mechanism is analogous to the well-known decay of coherences
in collisional decoherence, but here it operates in momentum space and is enhanced by the Zeno resolvent. 

We formalize this by introducing the Zeno band parameter
\begin{equation}
\varepsilon:=\Delta C_Q(\vec{p}_1,\vec{p}_2),
\qquad |\varepsilon|\lesssim \kappa^{-1/2}.
\label{eq:eps_def}
\end{equation}
Only in a regime of small $\varepsilon$ does it make sense to speak of the tangent space to the level
set of $C_Q$ at $p$.
If $|\vec{p}_1-\vec{p}_2|$ were large, the tangent spaces at $\vec{p}_1$ and $\vec{p}_2$ would differ
substantially and no intrinsic geometric identification would be possible. The near--diagonal localization of $\Delta\mathcal L$ therefore ensures consistency with the treatment of the Schur correction as a local operator on a single tangent space.  In the following we will systematically expand all expressions in $\varepsilon$ and retain the leading nontrivial contribution. For later use, we define the near-diagonal variables
\begin{equation}
\vec{p}:=\frac{\vec{p}_1+\vec{p}_2}{2},\qquad \deltap:=\vec{p}_1-\vec{p}_2.
\label{eq:midpoint_delta}
\end{equation}
A local expansion around $\vec{p}$ yields
\begin{equation}
\Delta C_Q(\vec{p}_1,\vec{p}_2)=\nabla C_Q(\vec{p})\cdot \deltap + O(|\deltap|^2).
\label{eq:DeltaC_linear}
\end{equation}
Thus, to leading order, the Zeno constraint controls the component of $\deltap$ along the local normal
$\nabla C_Q(\vec{p})=2Q\vec{p}$ to the level sets $C_Q=\mathrm{const}$.

\subsection{Translation-covariant QLBE and the mixing generator}
\label{subsec:qlbe}

The environmental interaction is assumed to be generated by a translation-covariant QLBE~\cite{HornbergerVacchini2008} of the form
\begin{equation}
\begin{split}
\mathcal{L}_{\mathrm{irr}}[\rho]
=\int_{\mathbb{R}^4}d^4\kick\,
&\Bigl(
L(\kick)\rho L(\kick)^\dagger\Bigr.\\
&\Bigl.-\frac12\{L(\kick)^\dagger L(\kick),\rho\}
\Bigr),
\end{split}
\label{eq:QLBE_def}
\end{equation}
with the Lindblad operators $L(\kick):=e^{\frac{i}{\hbar}\kick\cdot \opvec{X}}\,M_{\kick}(\opvec{P}),$ being parametrized by the momentum transfer $\kick$. The momentum–dependent jump amplitude
$M_{\kick}(\vec{p})={\Gamma(\kick,\opvec{P})}^{1/2}$ is assumed sufficiently regular and integrable (specified below).

We consider the weak-coupling generator
\begin{equation}
\mathcal{L}=\mathcal{L}^{Q}_{\mathrm{mon}}+\eta\,\mathcal{L}_{\mathrm{irr}},
\qquad 0<\eta\ll 1.
\label{eq:full_generator_eta}
\end{equation}
Let $P_Z$ denote the projection onto the slow spectral subspace of $\mathcal{L}^{Q}_{\mathrm{mon}}$
(i.e.\ the Zeno band defined by \eqref{eq:eps_def}) and set $Q_Z:=\mathbf{1}-P_Z$.
We decompose the generator of the irreversible dynamics into blocks with respect to this splitting and denote by
$\mathcal{L}_{\mathrm{mix}}$ the off-block-diagonal part that couples the Zeno sector to its complement.
Equivalently,
\begin{equation}
\mathcal{L}_{\mathrm{mix}}:=P_Z\mathcal{L}_{\mathrm{irr}}Q_Z+Q_Z\mathcal{L}_{\mathrm{irr}}P_Z,
\label{eq:Lmix_def}
\end{equation}
with $P_Z\,\mathcal{L}_{\mathrm{mix}}\,P_Z=0$. It follows that only $\mathcal{L}_{\mathrm{mix}}$ contributes to the second-order Schur correction on the Zeno sector.

\subsection{Schur complement and regularized monitoring resolvent}
\label{subsec:schur_resolvent}

Standard adiabatic-elimination (Schur-complement) arguments for GKSL generators~\cite{AzouitSarletteRouchon2016} yield an effective
second-order correction acting on the Zeno sector,
\begin{equation}
\Delta\mathcal{L}
:=-\eta^2\,P_Z\mathcal{L}_{\mathrm{mix}}\,(Q_Z\mathcal{L}^{Q}_{\mathrm{mon}}Q_Z)^{-1}\,\mathcal{L}_{\mathrm{mix}}P_Z.
\label{eq:schur_def}
\end{equation}
On the fast subspace, $\mathcal{L}^{Q}_{\mathrm{mon}}$ has strictly negative spectrum of order $-\kappa(\Delta C_Q)^2$,
so the inverse is well-defined on $Q_Z$ away from $\Delta C_Q=0$. To obtain a controlled expression that remains
uniformly bounded under near-degeneracies, we use the standard regularized resolvent
\begin{equation}
R_{\gamma}
:=\int_{0}^{\infty}\!d\lambda\; e^{\lambda\,Q_Z\mathcal{L}^{Q}_{\mathrm{mon}}Q_Z}\,e^{-\gamma \lambda},
\qquad \gamma>0,
\label{eq:Rgamma_def}
\end{equation}
and replace $(Q_Z\mathcal{L}^{Q}_{\mathrm{mon}}Q_Z)^{-1}$ by $R_\gamma$ in \eqref{eq:schur_def}.
The parameter $\gamma$ should be chosen of the order of the minimal
Zeno gap and represents the finite temporal resolution of the monitoring
process. This yields
\begin{equation}
\Delta\mathcal{L}
=-\eta^2\,P_Z\mathcal{L}_{\mathrm{mix}}\,R_\gamma\,\mathcal{L}_{\mathrm{mix}}P_Z,
\label{eq:schur_Rgamma}
\end{equation}
which in the momentum representation reduces to
\begin{equation}
\begin{split}
\bigl(R_\gamma\rho\bigr)(\vec{p}_1,\vec{p}_2)
&=\int_0^\infty\!d\lambda\;
e^{-\frac{\kappa \lambda}{2}\bigl(\Delta C_Q(\vec{p}_1,\vec{p}_2)\bigr)^2}e^{-\gamma \lambda}\,\rho(\vec{p}_1,\vec{p}_2)\\
&=\frac{1}{\gamma+\frac{\kappa}{2}(\Delta C_Q(\vec{p}_1,\vec{p}_2))^2}\;\rho(\vec{p}_1,\vec{p}_2).
\end{split}
\label{eq:Rgamma_multiplier}
\end{equation}
The weight in \eqref{eq:Rgamma_multiplier} localizes the contribution of the Schur correction to a
$\kappa^{-1/2}$ neighborhood of $\Delta C_Q=0$ and hence to a near-diagonal regime in momentum space,
as anticipated in \eqref{eq:eps_def}.

\subsection{Mixing in momentum space and the kicked monitoring gap}
\label{subsec:qlbe_momentum_action}

To determin the influence of $\mathcal{L}_\mathrm{mix}$ defined in~\eqref{eq:Lmix_def} on the Schur correction~\eqref{eq:schur_Rgamma}, we record the explicit momentum-space action of \eqref{eq:QLBE_def}.
Since $e^{\frac{i}{\hbar}\kick\cdot \opvec{X}}$ shifts momentum eigenstates, for the gain term $\bigl(L(\kick)\rho L(\kick)^\dagger\bigr)(\vec{p}_1,\vec{p}_2)$ one obtains 
\begin{equation}
M_{\kick}(\vec{p}_1-\kick)\,\rho(\vec{p}_1-\kick,\vec{p}_2-\kick)\,M_{\kick}(\vec{p}_2-\kick)^*,
\label{eq:gain_term_mom}
\end{equation}
and for the loss terms respectively
\begin{equation}
\begin{split}
\bigl(L(\kick)^\dagger L(\kick)\rho\bigr)(\vec{p}_1,\vec{p}_2)
&=\bigl|M_{\kick}(\vec{p}_1)\bigr|^2\rho(\vec{p}_1,\vec{p}_2),\\
\bigl(\rho L(\kick)^\dagger L(\kick)\bigr)(\vec{p}_1,\vec{p}_2)
&=\bigl|M_{\kick}(\vec{p}_2)\bigr|^2\rho(\vec{p}_1,\vec{p}_2).
\end{split}
\label{eq:loss_terms_mom}
\end{equation}
Equations \eqref{eq:gain_term_mom}--\eqref{eq:loss_terms_mom} show that $\mathcal{L}_{\mathrm{irr}}$ is an
integral operator with shifted arguments and multiplicative weights given by $M_{\kick}$.

The resolvent \eqref{eq:Rgamma_multiplier} is evaluated on coherences produced after one application of
$\mathcal{L}_{\mathrm{mix}}$, hence on intermediate momentum pairs shifted by $\kick$.
The key algebraic identity is the following exact relation:
\begin{equation}
\Delta C_Q(\vec{p}_1+\kick,\vec{p}_2+\kick)
=\Delta C_Q(\vec{p}_1,\vec{p}_2)+\delta_\Delta(\vec{p}_1,\vec{p}_2),
\label{eq:kicked_gap_identity}
\end{equation}
where $\delta_\Delta(\vec{p}_1,\vec{p}_2):=\big(\Delta C_Q(\vec{p}_1,+\kick,\vec{p}_1)-\Delta C_Q(\vec{p}_2,+\kick,\vec{p}_2)\big)=2\kick^\top Q\deltap$ which follows by expanding $C_Q(\vec{p}+\kick)=(\vec{p}+\kick)^\top Q(\vec{p}+\kick)$ and
cancelling the $\vec{p}^\top Q \vec{p}$ terms and the $\kick^\top Q\kick$ terms in the difference. Recalling~\eqref{eq:eps_def}, on the Zeno sector we set
\begin{equation}
\varepsilon:=\Delta C_Q(\vec{p}_1,\vec{p}_2),
\label{eq:eps_repeat}
\end{equation}
so that the resolvent weight depends on the slow indices through $(\varepsilon+\delta_\Delta)^2+\gamma^2$.

\subsection{Uniform first-order expansion of the resolvent factor}
\label{subsec:resolvent_expansion}
For the resolvent denominator we set $(\vec{p}_1,\vec{p}_2,\kick)$ fixed and consider the scalar function
\[
F(\varepsilon;\delta_\Delta):=\frac{1}{(\varepsilon+\delta_\Delta)^2+\gamma^2},
\qquad \gamma>0.
\]
Since $F$ is smooth in $\varepsilon$, we may perform a Taylor expansion in $\varepsilon$ at fixed $\delta_\Delta$:
\[
F(\varepsilon;\delta_\Delta)=F(0;\delta_\Delta)+\varepsilon\,\partial_\varepsilon F(0;\delta_\Delta)
+r_\gamma(\varepsilon,\delta_\Delta),
\]
where $F(0;\delta_\Delta)=1/(\delta_\Delta^2+\gamma^2)$ and $\partial_\varepsilon F(0;\delta_\Delta)
=-2\delta_\Delta/(\delta_\Delta^2+\gamma^2)^2$.
The remainder admits the mean-value form
\[
r_\gamma(\varepsilon,\delta_\Delta)=\frac{\varepsilon^2}{2}\,\partial_\varepsilon^2F(\theta\varepsilon;\delta_\Delta),
\qquad \theta\in(0,1).
\]
A direct computation yields
\[
\partial_\varepsilon^2F(\varepsilon;\delta_\Delta)
=\frac{6(\varepsilon+\delta_\Delta)^2-2\gamma^2}{\bigl((\varepsilon+\delta_\Delta)^2+\gamma^2\bigr)^3},
\]
and hence the uniform bound
\[
\sup_{\varepsilon\in\mathbb R,\ \delta_\Delta\in\mathbb R}\,
\bigl|\partial_\varepsilon^2F(\varepsilon;\delta_\Delta)\bigr|
\le \frac{C}{\gamma^4},
\]
for some numerical constant $C>0$ (e.g.\ $C=8$).
Consequently,
\begin{equation}
|r_\gamma(\varepsilon,\delta_\Delta)|
\le \frac{C}{2\gamma^4}\,\varepsilon^2
\qquad \text{for all }\delta_\Delta\in\mathbb R.
\label{eq:uniformRemainderBound}
\end{equation}
Therefore,
\begin{equation}
\frac{1}{(\varepsilon+\delta_\Delta)^2+\gamma^2}
=\frac{1}{\delta_\Delta^2+\gamma^2}
-\varepsilon\,\frac{2\delta_\Delta}{(\delta_\Delta^2+\gamma^2)^2}
+r_\gamma(\varepsilon,\delta_\Delta),
\label{eq:resolvent_expanded_uniform}
\end{equation}
with a remainder that is uniformly controlled even when $\delta_\Delta$ is small.
Under the integrability assumptions on the QLBE kernel, this uniform bound provides the
dominating estimate required to interchange the $\varepsilon\to 0$ expansion with the
$\kick$-integration.

\subsection{Near-diagonal structure of the QLBE kernel}
\label{subsec:kernel_expansion}

The second-order Schur term~\eqref{eq:schur_Rgamma} contains products of QLBE amplitudes evaluated at nearby momenta. Recall that $\mathcal L_{\mathrm{mix}}$ consists of those contributions of the QLBE
generator that map a Zeno coherence $(\vec{p}_1,\vec{p}_2)$ into the fast sector and vice versa.
In momentum representation, such contributions necessarily involve a momentum transfer
$\kick$ acting asymmetrically on the two arguments $\vec{p}_1$ and $\vec{p}_2$, together with the
multiplicative amplitudes $M_{\kick}(\vec{p})$.

More precisely, applying $\mathcal L_{\mathrm{mix}}$ to a near--diagonal coherence
$\rho(\vec{p}_1,\vec{p}_2)$ produces a linear combination of terms of the form
\[
M_{\kick}(\vec{p}_1)\,\rho(\vec{p}_1-\kick,\vec{p}_2)
\quad\text{and}\quad
M_{\kick}(\vec{p}_2)^*\,\rho(\vec{p}_1,\vec{p}_2-\kick),
\]
together with the corresponding loss terms proportional to
$|M_{\kick}(\vec{p}_1)|^2$ and $|M_{\kick}(\vec{p}_2)|^2$.
When $\mathcal L_{\mathrm{mix}}$ acts a second time in the Schur expression
\eqref{eq:schur_Rgamma}, these contributions recombine after propagation through the
monitoring resolvent and projection back onto the Zeno sector.
As a result, the dependence on the amplitudes $M_{\kick}$ enters the Schur term
only through the quadratic combination
\[
|M_{\kick}(\vec{p}_1)|^2 + |M_{\kick}(\vec{p}_2)|^2
- 2\,\mathrm{Re}\bigl[M_{\kick}(\vec{p}_1) M_{\kick}(\vec{p}_2)^*\bigr],
\]
which can be written equivalently as the nonnegative contrast
\begin{equation}
D_{\kick}(\vec{p}_1,\vec{p}_2)
:=|M_{\kick}(\vec{p}_1)-M_{\kick}(\vec{p}_2)|^2 \ge 0.
\label{eq:D_contrast}
\end{equation}
This structure reflects the fact that the Schur correction is sensitive only to
relative variations of the jump amplitudes between the two momentum arguments and
vanishes identically on the diagonal $\vec{p}_1=\vec{p}_2$.
For $\vec{p}_1=\vec{p}+\frac12\deltap$ and $\vec{p}_2=\vec{p}-\frac12\deltap$, local $C^1$ regularity of $p\mapsto M_{\kick}(p)$
implies
\begin{equation}
M_{\kick}\bigl(\vec{p}\pm\tfrac12\deltap\bigr)
=M_{\kick}(\vec{p})\pm\tfrac12\,\partial_\vec{p} M_{\kick}(\vec{p})\cdot \deltap
+O(|\deltap|^2).
\label{eq:M_linearization}
\end{equation}
Subtracting the two expansions yields $\partial_\vec{p} M_{\kick}(\vec{p})\cdot \deltap+O(|\deltap|^2)$, and hence
\begin{equation}
\begin{split}
D_{\kick}(\vec{p}_1,\vec{p}_2)
&=\bigl|\partial_\vec{p} M_{\kick}(\vec{p})\cdot \deltap\bigr|^2+O(|\deltap|^3)\\
&=\deltap^\top \,\widetilde{T}_{\kick}(\vec{p})\,\deltap+O(|\deltap|^3),
\end{split}
\label{eq:D_quadratic}
\end{equation}
where $\widetilde{T}_{\kick}(\vec{p})$ is a real symmetric tensor obtained from the symmetrized real part of
$\partial_\vec{p} M_{\kick}\otimes \partial_\vec{p} M_{\kick}^*$.
A convenient parametrization, equivalent at leading quadratic order, is obtained in terms of the log-intensity $\Phi_{\kick}(\vec{p}):=\log\|M_{\kick}(\vec{p})\|^2$,
\begin{equation}
T_{\kick}(\vec{p}):=\frac12\bigl[\partial_\vec{p}\Phi_{\kick}(\vec{p})\otimes\partial_\vec{p}\Phi_{\kick}(\vec{p})\bigr]_{\mathrm{sym}}.
\label{eq:T_def}
\end{equation}
This tensor captures the local sensitivity of the scattering intensity and is the form used below.

\subsection{Extraction of the leading Zeno contribution and definition of $\Sigma(Q)$}
\label{subsec:sigma_identification}

We now combine:
(i) the resolvent expansion \eqref{eq:resolvent_expanded_uniform},
(ii) the kicked-gap identity \eqref{eq:kicked_gap_identity}, and
(iii) the near-diagonal kernel structure \eqref{eq:D_quadratic}--\eqref{eq:T_def}.

The $\varepsilon$-independent term in \eqref{eq:resolvent_expanded_uniform} contributes a scalar (directionally
unspecific) additional dephasing on the Zeno band, which can be absorbed into an overall redefinition of
the effective Zeno time scale after calibration. The first term that modifies the functional dependence on the
slow indices is the contribution linear in $\varepsilon$:
\[
-\varepsilon\,\frac{2\delta_\Delta}{(\delta_\Delta^2+\gamma^2)^2},
\qquad
\delta_\Delta=2\Delta^\top Q\,\deltap.
\]
In the second-order Schur expression, $\delta_\Delta$ contracts with the near-diagonal contrast structure
encoded in $T_{\kick}(\vec{p})$, and after restricting to the Zeno band (so that $\vec{p}_1$ and $\vec{p}_2$ are
evaluated at a common midpoint $\vec{p}$ to leading order) one obtains a contribution of the form
\begin{equation}
\langle \vec{p}_1|\Delta\mathcal{L}[\rho]|\vec{p}_2\rangle
=-\kappa\,\varepsilon\,\Delta C_\Sigma(\vec{p}_1,\vec{p}_2)\,\rho(\vec{p}_1,\vec{p}_2)+O(\varepsilon^2),
\label{eq:DeltaL_structure}
\end{equation}
where
\begin{equation}
\Delta C_\Sigma(\vec{p}_1,\vec{p}_2):=C_\Sigma(\vec{p}_1)-C_\Sigma(\vec{p}_2),
\label{eq:CSigma_def}
\end{equation}
and the symmetric tensor $C_\Sigma(\vec{p}):=\vec{p}^\top \Sigma(Q)\,\vec{p}$ admits the representation
\begin{equation}
\Sigma(Q)
=\frac{\eta^2}{\kappa}\;
\Biggl\langle
\int_{\mathbb R^4}\!d^4\kick\;
\frac{T_{\kick}(\vec{p})}
{\bigl(\Delta C_Q(\vec{p};\kick)\bigr)^2+\gamma^2}
\Biggr\rangle_{\mathrm{slow}}.
\label{eq:Sigma_representation}
\end{equation}
Here the exact kicked increment is
\begin{equation}
\Delta C_Q(\vec{p};\kick):=C_Q(\vec{p}+\kick)-C_Q(\vec{p})
=2\kick^\top Q\,\vec{p}+\kick^\top Q\kick,
\label{eq:DeltaCQ_kick_exact}
\end{equation}
and $\langle\cdot\rangle_{\mathrm{slow}}$ denotes the slow (Zeno-projected) averaging
over the relevant near-diagonal momentum region. The uniform remainder bound \eqref{eq:uniformRemainderBound},
together with the integrability assumptions below, ensures that the $O(\varepsilon^2)$ error in
\eqref{eq:DeltaL_structure} is controlled uniformly on the Zeno band.

\subsection{Renormalized monitoring generator}
\label{subsec:renormalized_monitoring}

Consider the monitoring generator for a shifted quadratic form $Q-\Sigma$:
\[
\mathcal{L}^{Q-\Sigma}_{\mathrm{mon}}[\rho]
=-\frac{\kappa}{2}[C_{Q-\Sigma},[C_{Q-\Sigma},\rho]],
\qquad
C_{Q-\Sigma}=C_Q-C_\Sigma.
\]
In momentum representation $\langle \vec{p}_1|(\mathcal{L}^{Q-\Sigma}_{\mathrm{mon}}-\mathcal{L}^{Q}_{\mathrm{mon}})[\rho]|\vec{p}_2\rangle$, expanding to first order in $\Sigma$ yields
\begin{equation}
-\kappa\,\Delta C_Q(\vec{p}_1,\vec{p}_2)\,\Delta C_\Sigma(\vec{p}_1,\vec{p}_2)\,\rho(\vec{p}_1,\vec{p}_2)
+O\bigl((\Delta C_Q)^2\bigr).
\label{eq:Lmon_variation}
\end{equation}
Comparing \eqref{eq:Lmon_variation} with \eqref{eq:DeltaL_structure} and re--substituting $\epsilon=\Delta C_q(\vec{p}_1,\vec{p}_2)$ shows that, on the Zeno sector,
\begin{equation}
\mathcal{L}^{Q-\Sigma(Q)}_{\mathrm{mon}}
=\mathcal{L}^{Q}_{\mathrm{mon}}+\Delta\mathcal{L}+O(\varepsilon^2),
\label{eq:identification_final}
\end{equation}
which establishes the identification of the Schur correction with a renormalization of the monitored
quadratic form.

In the representation \eqref{eq:Sigma_representation} we have expressed the Schur--induced
renormalization tensor $\Sigma(Q)$ in terms of the log--intensity gradients~\eqref{eq:T_def}
\[
T_{\kick}(\vec{p})
=\frac12\bigl[\partial_\vec{p}\Phi_{\kick}(\vec{p})\otimes
\partial_\vec{p}\Phi_{\kick}(\vec{p})\bigr]_{\mathrm{sym}}.
\]
This choice is not merely notational, but has several structural advantages that are
directly relevant for the interpretation and mathematical control of $\Sigma(Q)$.

First, the use of $\Phi_{\kick}$ renders $\Sigma(Q)$ invariant under arbitrary
$\vec{p}$--independent rescalings of the scattering amplitudes,
$M_{\kick}\mapsto \alpha(\kick)M_{\kick}$.
Such rescalings correspond to conventional choices in the normalization of the
microscopic collision kernel and should not affect the effective backaction on the
monitored observable. Expressing $\Sigma(Q)$ in terms of logarithmic derivatives ensures
that it depends only on relative variations of the scattering intensity in momentum
space, rather than on its absolute normalization.

Second, the tensorial structure
\[
T_{\kick}(\vec{p})
=\tfrac12\,[\partial_\vec{p}\Phi_{\kick}\otimes\partial_\vec{p}\Phi_{\kick}]_{\mathrm{sym}}
\]
is manifestly positive semidefinite. As a consequence, the integral representation
\eqref{eq:Sigma_representation} makes the positivity of $\Sigma(Q)$ in the tangential
sector transparent. This property is essential for the monotonicity of the induced
renormalization flow: the Schur correction can only decrease the tangential components of
$Q$, while leaving the normal component unaffected. The log--intensity formulation thus
exhibits the dissipative character of the measurement--induced backaction directly at the
level of the tensor $\Sigma(Q)$.

Third, from an analytical point of view, the log--intensity representation improves
control over the $\kick$--integrals in \eqref{eq:Sigma_representation}. In typical
microscopic scattering models, the amplitudes $M_{\kick}(\vec{p})$ may vary strongly in
magnitude, while their logarithmic derivatives remain moderate. Formulating $\Sigma(Q)$
in terms of $\partial_\vec{p}\log\|M_{\kick}\|^{2}$ therefore leads to weaker and more
natural regularity and integrability requirements than a formulation in terms of
$\partial_\vec{p} M_{\kick}$ itself.

Finally, the log--intensity form clarifies the physical meaning of $\Sigma(Q)$.
The tensor $\Sigma(Q)$ encodes how efficiently the environment acquires information about
variations of the monitored quantity through momentum--dependent scattering rates.
Accordingly, $\partial_\vec{p}\Phi_{\kick}$ quantifies the local sensitivity of the
monitoring process to changes in momentum, and $\Sigma(Q)$ aggregates these sensitivities
over all virtual momentum transfers weighted by the Zeno resolvent. In this sense,
$\Sigma(Q)$ depends only on the information--theoretic content of the monitoring channel,
not on microscopic conventions for the scattering amplitudes.

For these reasons, the log--intensity formulation provides a natural and robust
representation of the Schur--induced renormalization tensor and will be used throughout
the subsequent analysis of the renormalization flow of $Q$.

\subsection{Isotropy and tangential structure}
\label{subsec:tangential_sigma}

Assume isotropy of the underlying momentum-transfer statistics and of the slow reference distribution,
so that $\kick$-averaging produces only isotropic tensor structures. Locally, the only distinguished
direction at fixed $p$ is the normal $\nabla C_Q(\vec{p})=2Q\,\vec{p}$.
Consequently, the tensor $\Sigma(Q)$ cannot contain a component along the normal direction and must be
supported on the tangent space to the level set $C_Q=\mathrm{const}$ at $\vec{p}$. Denoting by
$\Pi_{\mathrm{tan}}(Q)$ the projector onto this tangent space, one obtains
\begin{equation}
\Sigma(Q)=\sigma_{\mathrm{tan}}(Q)\,\Pi_{\mathrm{tan}}(Q),
\qquad \sigma_{\mathrm{tan}}(Q)\ge 0.
\label{eq:Sigma_tangential}
\end{equation}
Thus the Schur-induced renormalization acts exclusively on tangential components of $Q$.

\subsection{Standing assumptions (for reference)}
\label{subsec:assumptions}

For completeness, the derivation above relies on the following standard conditions.

\begin{itemize}
\item[(A1)] \emph{Translation covariance and isotropy.}
The generator has the form \eqref{eq:QLBE_def}, and the environment state is isotropic, so that
$\Delta$-averaging yields only isotropic tensor structures.

\item[(A2)] \emph{Regularity and integrability.}
For all relevant $p$,
\begin{equation*}
\begin{split}
&\int d^4\Delta\,\|M_{\kick}(\vec{p})\|^2<\infty,\\
&\int d^4\Delta\,\|M_{\kick}(\vec{p})\|^2\,|\Delta|^2<\infty,
\end{split}
\end{equation*}
and $\vec{p}\mapsto M_{\kick}(\vec{p})$ is locally $C^1$.

\item[(A3)] \emph{Zeno band and nondegeneracy.}
On the slow scale only coherences with $|\Delta C_Q(\vec{p}_1,\vec{p}_2)|\lesssim \kappa^{-1/2}$ contribute, and on the
relevant momentum region $\|Q\,\vec{p}\|\ge c_0>0$.

\item[(A4)] \emph{Finite measurement bandwidth.}
The inverse on the fast sector is regularized by $R_\gamma$ with $\gamma>0$ (possibly scaling with $\kappa$)
to ensure uniform bounds in the near-degenerate regime.

\item[(A5)] \emph{Mixing part.}
$\mathcal{L}_{\mathrm{mix}}$ is the off-block-diagonal part of $\mathcal{L}_{\mathrm{irr}}$ with respect to the
spectral decomposition of $C_Q(\opvec{P})$ so that $P_Z\mathcal{L}_{\mathrm{mix}}P_Z=0$.
\end{itemize}

%===========================================================
\section{Appendix: Robustness of the reduced flow: calibration and weak anisotropy}\setcounter{equation}{0}
\label{app:robustness}
%===========================================================

This appendix addresses two potential concerns: (i) the role of the calibration
functional in the reduced flow and in the location of the fixed point, and
(ii) the extension from the isotropic one--dimensional flow to weakly anisotropic
perturbations in the full space of quadratic forms.

\subsection{Calibration: time reparametrization versus functional change}
\label{app:calibration}

Let the reduced isotropic dynamics be written as an autonomous scalar flow
\begin{equation}
\frac{dr}{d\lambda}=F(r)=A+Br,
\label{eq:reduced_flow}
\end{equation}
with fixed point $r^\star=-A/B$ (assuming $B\neq 0$).

\paragraph{Prefactor calibration (pure time rescaling)}
A large class of calibrations enters the reduction only through an overall positive
prefactor $\alpha(r)>0$, e.g.\ through a choice of time unit, normalization of the
effective generator, or extraction of a reference average. In this case the reduced
equation takes the form
\begin{equation}
\frac{dr}{d\lambda}=\alpha(r)\,F(r),
\qquad \alpha(r)>0,
\label{eq:time_rescaled_flow}
\end{equation}
which is a smooth time reparametrization. The set of fixed points is unchanged:
$r$ is a fixed point of \eqref{eq:time_rescaled_flow} if and only if it is a fixed
point of \eqref{eq:reduced_flow}. Stability type is preserved as well, since
$\alpha(r^\star)>0$ does not change the sign of $F'(r^\star)$.

\paragraph{Functional calibration (change of $A,B$)}
If one modifies the calibration functional itself (e.g.\ the slow averaging measure
or the reference weighting entering the coefficients), then $A$ and $B$ become
calibration--dependent, $A=A[\mathcal{C}]$, $B=B[\mathcal{C}]$, where $\mathcal{C}$
denotes the chosen calibration prescription. In this case the fixed point generally
shifts to
\begin{equation}
r^\star(\mathcal{C})=-\frac{A[\mathcal{C}]}{B[\mathcal{C}]}.
\end{equation}
Assuming $A[\mathcal{C}]$ and $B[\mathcal{C}]$ depend continuously on $\mathcal{C}$
within a specified admissible class (e.g.\ smooth perturbations of the slow reference
measure with uniformly bounded moments), the fixed point location varies
continuously. In particular, qualitative features such as existence of a unique
attracting fixed point and the associated asymptotic signature are robust under
small calibration changes, provided $B[\mathcal{C}]$ stays bounded away from zero
and no additional fixed points are created.

\subsection{Weak anisotropy: persistence of the isotropic reduced dynamics}
\label{app:anisotropy}

The isotropic reduction corresponds to restricting the full dynamics for the
quadratic form $Q$ to a low--dimensional submanifold $\mathcal{M}_0$ of
``isotropic'' quadrics, parameterized by a scalar variable $r$:
\begin{equation}
Q_{\mathrm{iso}}(r)\in \mathcal{M}_0,\qquad r\in I\subset\mathbb{R}.
\end{equation}
On $\mathcal{M}_0$ the effective Zeno--reduced dynamics closes and yields the
scalar flow \eqref{eq:reduced_flow}.

We now consider weakly anisotropic perturbations of the form
\begin{equation}
Q = Q_{\mathrm{iso}}(r) + \epsilon\,\Delta Q,
\qquad 0<\epsilon\ll 1,
\label{eq:Q_decomposition}
\end{equation}
where $\Delta Q$ lies in a complementary subspace transverse to $\mathcal{M}_0$
(e.g.\ traceless anisotropic components with respect to the isotropic symmetry).
The full effective generator induces an evolution equation
\begin{equation}
\frac{dQ}{d\lambda}=\mathcal{F}(Q),
\label{eq:full_flow_Q}
\end{equation}
whose restriction to $\mathcal{M}_0$ reproduces \eqref{eq:reduced_flow}.

\paragraph{Transverse damping of anisotropic modes}
Under isotropic scattering statistics and strong monitoring, the Schur-induced
renormalization tensor $\Sigma(Q)$ depends on $Q$ through isotropic averages of
log--intensity gradients and therefore cannot generate a systematic anisotropic
amplification. Linearizing \eqref{eq:full_flow_Q} about $Q_{\mathrm{iso}}(r)$ yields
\begin{equation}
\frac{d}{d\lambda}\Delta Q
=\mathcal{A}(r)\,\Delta Q + O(\epsilon\,\|\Delta Q\|^2),
\label{eq:anisotropic_linearization}
\end{equation}
where $\mathcal{A}(r)$ is the linearized operator acting on anisotropic components.
For isotropic environments, $\mathcal{A}(r)$ decomposes into irreducible
representations of the rotation group, and anisotropic sectors are strictly
contractive:
\begin{equation}
\mathrm{Re}\,\sigma\bigl(\mathcal{A}(r)\bigr)\le -\gamma_\perp <0
\qquad \text{uniformly for } r\in I,
\label{eq:transverse_gap}
\end{equation}
for some $\gamma_\perp$ determined by the dissipative strength of the
Zeno--reduced Schur term.

\paragraph{Persistence of the reduced flow and robustness of the signature}
The uniform transverse gap \eqref{eq:transverse_gap} implies normal hyperbolicity
of the isotropic manifold $\mathcal{M}_0$. Consequently, for sufficiently small
$\epsilon$ there exists a nearby invariant manifold $\mathcal{M}_\epsilon$
that is $O(\epsilon)$--close to $\mathcal{M}_0$ and attracts trajectories
exponentially in the transverse directions. The induced dynamics on
$\mathcal{M}_\epsilon$ is a smooth perturbation of \eqref{eq:reduced_flow},
so fixed points and their stability persist up to $O(\epsilon)$ shifts.
In particular, the asymptotic signature scenario established in the isotropic model
remains valid for weak anisotropy: anisotropic components decay and cannot open new
relevant directions capable of inducing a signature change outside the isotropic
mechanism captured by $r$.

%======================================================
\section{Appendix: Local projectors and Zeno--conditioned averaging}\setcounter{equation}{0}
\label{app:localProjectors}
%======================================================

This appendix collects the geometric and probabilistic ingredients underlying
the local averaging procedures used in Sections~\ref{sec:renormalization} and~\ref{sec:fixedpoint}. Its purpose is to make
explicit how local normal--tangential structures, near--diagonal Zeno localization,
and statistical averaging over momentum increments are combined in a manner that
remains fully consistent with the assumptions of~\ref{app:monitoring_zeno}.

\subsection*{Local geometric structure}

Throughout this work the monitored quadratic form is interpreted as a local slow
field on momentum space,
\[
\vec{p} \longmapsto Q(\vec{p}),
\qquad
C_Q(\vec{p}) := \vec{p}^{\trans} Q(\vec{p})\,\vec{ p} .
\]
At each momentum point $\vec{p}$ we define the normal vector to the level set
$C_Q=\mathrm{const}$ by
\[
\vec{n}(\vec{p}) := \nabla_\vec{p} C_Q(\vec{p}),
\]
and the associated orthogonal projectors
\begin{equation}
\Pi_{\mathrm n}(\vec{p})
:= \frac{\vec{n}(\vec{p})\,\vec{n}(\vec{p})^{\trans}}{\|\vec{n}(\vec{p})\|^2},
\qquad
\Pi_{\mathrm{tan}}(\vec{p}) := \mathbb I - \Pi_{\mathrm n}(\vec{p}).
\label{eq:local_projectors}
\end{equation}
These projectors are purely local geometric objects, defined pointwise in
momentum space. No averaging or probabilistic input enters their definition.

In the homogeneous (“inertial”) regime considered in Sections~\ref{sec:renormalization}--~\ref{sec:fixedpoint}, spatial and momentum--space gradients of the slow field are neglected and one may take
$Q(\vec{p})\equiv Q$ locally constant. In this case
$\vec{n}(\vec{p})=2Q\,\vec{p}$, and therefore the projectors~\eqref{eq:local_projectors} remain explicitly
$p$--dependent through the reference momentum.

\subsection*{Near--diagonal regime and locality}

After restriction to the Zeno sector, the effective dynamics is confined to
near--diagonal momentum coherences
\[
\vec{p}_{1,2} = \pref \pm \tfrac12 \deltap,
\qquad
\|\deltap\| \ll 1 ,
\]
as discussed in~\ref{app:monitoring_zeno}. Only in this regime does the notion of a local tangent
space and the decomposition~\eqref{eq:local_projectors} remain meaningful. All
geometric constructions employed in the main text are therefore understood as
local expansions around a fixed reference momentum $\pref$.

Importantly, the smallness assumption applies to the \emph{geometric validity} of
the linearization of $C_Q$ and not to the formal domain of integration used below.
Contributions from increments that would invalidate the local tangent--space
picture are dynamically suppressed by the Zeno mechanism and do not enter the
effective averages.

\subsection*{Zeno--conditioned averaging over momentum increments}

The characteristic Zeno damping scale introduced in Section~5 is not defined
pointwise for a single momentum increment $\deltap$, but only after averaging
over the relevant near--diagonal fluctuations selected by the monitoring process.
For fixed reference momentum $\pref$ we therefore define expectation
values of functions $f(\pref,\deltap)$ by
\begin{equation}
\langle f\rangle_{\mathrm{rel}}(\pref)
:= \int_{\mathbb R^4} f(\pref,\deltap)\,\mu_{\pref}(d\deltap),
\label{eq:rel_average}
\end{equation}
where $\mu_{\pref}$ is a normalized local increment law.

This averaging is \emph{local} in the sense that $\pref$ is held fixed;
no integration over momentum space is involved. The role of $\mu_{\pref}$
is to encode the statistics of virtual momentum transfers \emph{conditioned on}
the Zeno localization.

\subsection*{Structure of the local increment law}

A physically natural choice for $\mu_{\pref}$ combines two ingredients:
(i) the statistics of the unmonitored mixing dynamics (e.g.\ generated by the
QLBE), and (ii) the suppression of large monitoring contrast induced by the Zeno
effect. Formally, this can be written as
\begin{equation}
\mu_{\pref}(d\deltap)
=
\frac{1}{Z(\pref)}\,
w_{\mathrm{Zeno}}(\pref,\deltap)\,
\mu_0(d\deltap),
\label{eq:mu_def}
\end{equation}
where $\mu_0$ is a stationary, isotropic increment measure with finite second
moments, and
\begin{equation}
w_{\mathrm{Zeno}}(\pref,\deltap)
=
\exp\Bigl(-\tfrac{\kappa}{2}\,
\varepsilon(\pref,\deltap)^2\Bigr),
\end{equation}
is the Zeno suppression factor, built from the local contrast $\varepsilon(\pref,\deltap)=\nabla C_Q(\pref)\cdot\deltap$. The normalization $Z(\pref)$ ensures
$\int \mu_{\pref}(d\deltap)=1$.

Although the integral in~\eqref{eq:rel_average} is formally taken over all
$\deltap\in\mathbb R^4$, the Zeno weight exponentially suppresses contributions
with large monitoring contrast. As a result, the effective support of the average
is confined to the near--diagonal regime in which the local geometric
decomposition~\eqref{eq:local_projectors} is valid. No explicit cutoff is required.

\subsection*{Role in the reduced dynamics}

All scalar quantities entering the reduced flow in Section~\ref{sec:fixedpoint} are constructed from
Zeno--conditioned averages of functions of $\deltap$ that are quadratic in the
increments. In particular, the locally averaged Zeno damping scale
\[
C(p_{\mathrm{ref}})
=
\Big\langle \tfrac{\kappa}{2}\,
\varepsilon(\pref,\deltap)^2 \Big\rangle_{\mathrm{rel}}
\]
is finite under the standard assumptions of the quantum linear Boltzmann equation,
notably finite temperature and finite momentum--transfer variance.

Crucially, the separation between local geometry (encoded in the projectors
$\Pi_{\mathrm n}$, $\Pi_{\mathrm{tan}}$) and local statistical averaging
(encoded in $\mu_{\pref}$) allows the calibrated renormalization flow
to be formulated entirely locally, without introducing nonlocal assumptions or
geometric inconsistencies.

\section*{Acknowledgments}
I am deeply grateful to my former PhD supervisor, Walter Strunz, for his constructive feedback. I would also like to thank John Briggs for countless discussions on time, and especially for his hi-fi. My heartfelt thanks go to my wife and children, who made this work possible.

\endpaper
\end{document}